\documentclass[10pt, a4paper, twocolumn]{article}
\usepackage[left=2cm, right=2cm, top=3cm, bottom=3cm]{geometry}
\usepackage[utf8]{inputenc}
\usepackage[english]{babel}

\usepackage{xcolor}
\definecolor{darkred}{RGB}{220, 0, 0}

\definecolor{bluer}{RGB}{50, 50, 150}
\definecolor{redr}{RGB}{220, 50, 50}
\definecolor{pinkr}{RGB}{200, 0, 100}

\usepackage{amsmath}
\usepackage{amsfonts}
\usepackage{dsfont}
\usepackage{braket}
\usepackage{mathtools}
\usepackage{lmodern}
\usepackage[
    colorlinks=true,
    citecolor=bluer,
    linkcolor=redr,
    urlcolor=pinkr,
]{hyperref}
\usepackage{multicol}
\usepackage{parskip}
\usepackage{cleveref}
\usepackage{stmaryrd}
\usepackage{mathrsfs}
\usepackage{changepage}
\usepackage[symbol,hang,flushmargin]{footmisc}
\usepackage{xurl}
\usepackage[explicit]{titlesec}
\usepackage{caption}
\usepackage{lipsum}
\usepackage{amsthm}
\usepackage{float}
\usepackage[shortlabels]{enumitem}
\usepackage{tikz}
\usepackage{amssymb}
\usepackage{float}
\usepackage{pgfplots}
\usepackage{cuted}

\usetikzlibrary{patterns}
\pgfplotsset{compat=newest}

\newcommand{\potato}[5]{
    \draw[thick, cyan!40!black, pattern=crosshatch dots, pattern color=cyan!50!black, shift={#1}, rotate=#2, xscale=#5] plot[smooth cycle] coordinates {(0,0) (1*0.15,2*0.15) (3*0.15,1*0.15) (2*0.15,-1*0.15) (1*0.15,-2*0.15)};
    \node at #3 {#4};
}

\pgfmathdeclarefunction{gauss}{3}{%
  \pgfmathparse{1/(#3*sqrt(2*pi))*exp(-((#1-#2)^2)/(2*#3^2))}%
}

\titleformat{\section}{\normalsize\bfseries}{\thesection.}{0.4em}{\MakeUppercase{#1}}
\titleformat{\subsection}{\normalsize\bfseries}{\thesubsection.}{0.4em}{#1}

\DeclareRobustCommand\ccdots{\mathinner{\mkern-4mu\cdotp\mkern-3mu\cdotp\mkern-3mu\cdotp\mkern-3mu}}

\usepackage[backend=bibtex, style=numeric-comp, sorting=none, maxbibnames=10, hyperref=true, url=false, doi=false]{biblatex}
\newbibmacro{string+url}[1]{\href{\thefield{url}}{#1}}
\DeclareFieldFormat*{title}{\usebibmacro{string+url}{#1}}
\DeclareFieldFormat*{journaltitle}{\mkbibitalic{#1}\addcomma}
\renewbibmacro{in:}{}
\AtEveryBibitem{%
  \clearfield{month}%
  \clearfield{pages}%
  \clearfield{note}%
  \clearfield{isbn}%
}

\DeclareFieldFormat[book,inproceedings,unpublished]{date}{(#1)}
\DeclareListFormat[book,inproceedings]{publisher}{\mkbibitalic{#1}\nopunct}
\DeclareFieldFormat[inproceedings]{booktitle}{\mkbibitalic{#1}\addcomma}
\DeclareFieldFormat[misc]{url}{\href{#1}{#1}\nopunct}
\DeclareFieldFormat[unpublished]{note}{\mkbibitalic{#1}}

\DeclareBibliographyDriver{misc}{%
    \usebibmacro{begentry}%
        \printfield{url}%
    \usebibmacro{finentry}
}

\DeclareBibliographyDriver{unpublished}{%
    \usebibmacro{begentry}%
        \printnames{author}\adddot\space%
        \printfield{title}\adddot\space%
        \printfield{url}\adddot\space%
        In preparation\space%
        \printdate\adddot%
    \usebibmacro{finentry}
}

\DeclareBibliographyDriver{software}{%
    \usebibmacro{begentry}%
        \printnames{author}\adddot\space%
        \printfield{title}\adddot\space%
        \printfield{url}\adddot\space%
        (\printdate)\adddot%
    \usebibmacro{finentry}
}

\usepackage{graphicx}
\graphicspath{{figures/}}
\usepackage{caption}

\usepackage{xspace}

\newcommand{\dd}{\ensuremath{\mathrm{d}}}
\DeclarePairedDelimiterXPP\tmpTr[1]{\mathrm{Tr}}{[}{]}{}{#1}
\newcommand{\Tr}{\tmpTr*}
\DeclarePairedDelimiterXPP\tmpE[1]{\mathbb{E}}{[}{]}{}{#1}
\newcommand{\E}{\tmpE*}
\DeclarePairedDelimiterXPP\tmpEhat[1]{\hat{\mathbb{E}}}{[}{]}{}{#1}
\newcommand{\Ehat}{\tmpEhat*}
\DeclarePairedDelimiterXPP\tmppr[1]{\mathbb{P}}{[}{]}{}{#1}
\newcommand{\pr}{\tmppr*}
\newcommand{\dt}{\ensuremath{\dd t}}
\newcommand{\dYt}{\ensuremath{\dd Y_t}}

\newcommand{\dWt}{\ensuremath{\dd W_t}}
\newcommand{\dNt}{\ensuremath{\dd N_t}}
\newcommand{\rhobar}{\overline{\rho}}
\newcommand{\Ld}{\ensuremath{L^\dag}}
\newcommand{\Lcal}{\mathcal{L}}

\newcommand{\Scal}{\mathcal{S}}

\newcommand{\Hcal}{\mathcal{H}}
\newcommand{\Pcal}{\mathcal{P}}
\newcommand{\Dcal}{\mathcal{D}}
\newcommand{\Mcal}{\mathcal{M}}
\newcommand{\Zcal}{\mathcal{Z}}
\newcommand{\Ccal}{\mathcal{C}}
\newcommand{\Kcal}{\mathcal{K}}

\newcommand{\nexp}{{n_\text{exp}}}
\newcommand{\nsubset}{{n_\text{subset}}}
\newcommand{\1}{\mathds{1}}
\newcommand{\btheta}{{\boldsymbol{\theta}}}

\newcommand{\PP}{\mathbb{P}}
\newcommand{\knowing}{\,|\,}

\begin{document}

\begin{strip}
    \begin{center}
        {\fontsize{14}{0}\textbf{
            Parameters estimation by fitting correlation\\
            functions of continuous quantum measurement
        }}
        \\\vspace{0.5cm}
        {\large
            Pierre Guilmin\textsuperscript{1,2,}\footnotemark, Pierre Rouchon\textsuperscript{2} and Antoine Tilloy\textsuperscript{2}
        }
        \\\vspace{0.2cm}
        {\normalsize
            \textsuperscript{1}\textit{Alice \& Bob, 49 Bd du Général Martial Valin, 75015 Paris, France}
            \\
            \textsuperscript{2}\textit{Laboratoire de Physique de l’École Normale Supérieure, Mines Paris,}
            \\
            \textit{Inria, CNRS, ENS-PSL, Centre Automatique et Systèmes (CAS),}
            \\
            \textit{Sorbonne Université, PSL Research University, Paris, France}
            \\\vspace{0.2cm}
            \today
            \\\vspace{0.2cm}
        }
    \end{center}

    \begin{adjustwidth}{2cm}{2cm}
        We propose a simple method to estimate the parameters of a continuously measured quantum system, by fitting correlation functions of the measured signal. We demonstrate the approach in simulation, both on toy examples and on a recent superconducting circuits experiment which proved particularly difficult to characterise using conventional methods. The idea is applicable to any system whose evolution is described by a jump or diffusive stochastic master equation. It allows the simultaneous estimation of many parameters, is practical for everyday use, is suitable for large Hilbert space dimensions, and takes into account experimental constraints such as detector imperfections and signal filtering and digitisation. Unlike existing methods, it also provides a direct way to understand how each parameter is estimated from the measured signal. This makes the approach interpretable, facilitates debugging, and enables validating the adequacy of a model with the observed data.
    \end{adjustwidth}

    \vspace{0.5cm}

\end{strip}

\footnotetext{\href{mailto:pierre.guilmin@alice-bob.com}{pierre.guilmin@alice-bob.com}}

\section{Introduction}

Characterising the dynamics of a quantum system is a fundamental task in experimental quantum physics. Typically, a series of hand-crafted measurements are performed in a precise order to estimate the value of each system parameter in turn. However, as quantum technologies mature, this approach is no longer efficient to characterise increasingly large and complex systems, such as in the recent demonstrations of quantum error correction embedding multiple superconducting qubits in a surface code \cite{krinner2022realizing,zhao2022realization,acharya2023suppressing,acharya2024quantum}. A major challenge to streamline the development of these technologies is to find characterisation methods that (i) facilitate the automated estimation of many parameters, and (ii) remain practical for everyday use in the laboratory. In this paper, we focus on quantum systems that are measured continuously. We propose a simple method to characterise such systems, by directly fitting the statistics of the measured signals with an explicit formula.

The evolution of a continuously measured quantum system is modelled by the stochastic master equation (SME) formalism. The SME describes the evolution of the quantum state and the corresponding signal $I_t$ measured by the observer at time $t$. However, the quantity $I_t$ is never available to the experimentalist, it is merely a mathematical object, and not a tangible, measurable quantity. The actual measured signal is processed by an acquisition chain consisting of amplifiers, filters and digitisation components. This chain converts the continuous-time signal $I_t$ into a discrete-time signal $\{I_0, \dots, I_N\}$. Each value $I_k$ is defined by integrating $I_t$ against the filter function $f_k$ of the acquisition chain for the $k$-th time bin: ${I_k=\int f_k(t) I_t\,\dt}$. The only quantity available to the experimentalist is this digitised measurement record $\{I_0, \dots, I_N\}$. It is crucial to take this filtering into account when the digitisation time is not negligible compared to the system timescales, as is often the case for superconducting circuits, for example.

\begin{figure*}[ht]
    \centering
    \includegraphics{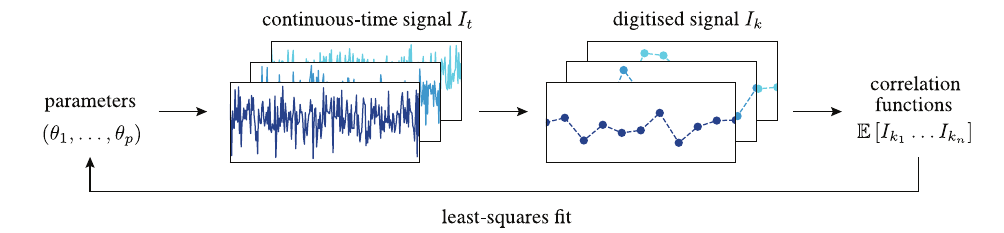}
    \caption{
        \textbf{Estimating parameters by fitting correlation functions.} A quantum system characterised by $p$ unknown parameters $(\theta_1,\dots,\theta_p)$ is continuously measured. Each realisation of the experiment (blue lines) yields a continuous-time signal $I_t$. In practice, this signal is filtered by an acquisition chain, resulting in a digitised signal $I_k=\int f_k(t)I_t\,\dt$, where $f_k$ is the filter function for the $k$-th time bin. The correlation functions of this digitised signal $\E{I_{k_1}\dots I_{k_n}}$ are estimated by averaging over multiple realisations of the experiment. The system parameters are determined by fitting these correlation functions using an exact theoretical formula.
    }
    \label{fig:main}
\end{figure*}

The statistics of this discrete signal can be estimated experimentally by performing the same experiment several times and averaging over the different realisations. These statistics include the mean of the signal $\E{I_k}$ at time bin $k$, the two-point correlation function $\E{I_{k_1}I_{k_2}}$ between bins $k_1$ and $k_2$, and more generally any $n$-point correlation functions $\E{I_{k_1}\dots I_{k_n}}$. Recently, an exact formula \cite{tilloy2018exact,guilmin2023correlation} has been derived to compute these quantities directly from the SME and the initial state of the system. In this paper, we propose to use this formula to estimate the system parameters by fitting the experimental estimates of the correlation functions. \Cref{fig:main} illustrates the method.

There are two other main approaches to estimating parameters from continuous measurement data. The first is based on Bayesian inference, the system parameters are determined by maximising the likelihood of the observed data \cite{mabuchi1996dynamical,gambetta2001state,chase2009single,negretti2013estimation,gammelmark2013bayesian,six2015parameter,ralph2017multiparameter,bompais2022parameter,clausen2024online}. These methods make optimal use of the available information, but the numerical cost is prohibitive for estimating many parameters, and they do not take into account signal filtering and digitisation. The second more recent approach is to train a machine learning model \cite{greplova2017quantum,krastanov2020unboxing,khanahmadi2021time,genois2021quantum,rinaldi2024parameter,tucker2024hamiltonian}, either in a supervised setting to learn the inverse mapping from observed data to parameters, or alternatively in an unsupervised setting to learn the dynamics using a model with a physics-motivated inductive bias. These methods are robust to noise and can also model non-Markovian dynamics, but they require a large training set for each new problem. Importantly, both the Bayesian and the learning approaches suffer from a lack of interpretability, which limits their usefulness for everyday applications. It is difficult to develop an intuition about how the data is being used to fit each parameter, which complicates the debugging process and reduces confidence in the estimated parameters.

In contrast, the philosophy of the proposed method is to directly fit the observed experimental data with an explicit formula, and without trying to reconstruct intermediate quantities, such as the quantum state. This characterisation scheme does not make optimal use of the available information, but it offers several advantages. First, it is \emph{general}: it applies to any quantum system modelled by a jump and/or diffusive SME, and takes into account common detector imperfections (efficiency and dark count rate), as well as arbitrary signal filtering and digitisation. Second, it is \emph{simple}: the experimental processing is straightforward and the fitting procedure is easy to implement numerically. Third, it is \emph{efficient}: it allows for the simultaneous estimation of multiple parameters, and is suitable for large Hilbert space dimensions. And fourth, it is \emph{interpretable}: an experimentalist can develop an intuition about how each parameter uniquely impacts the fitted correlation functions, and the goodness of fit can be asserted to validate the adequacy of the model with the observed data. It is therefore suitable for everyday use in the laboratory, as a new tool for characterising quantum systems. These advantages also make the approach an interesting choice for automating the characterisation phase. Recently, a similar idea based on polyspectra (Fourier transform of higher-order moments), has been successfully applied in the field of spin noise spectroscopy, quantum transport and semiconductor quantum dots \cite{hagele2018higher,sifft2021quantum,sifft2023random,sifft2024quantum}.

This paper is structured as follows. In \cref{sec:sme} we recall the SME formalism for continuous jump and diffusive measurements by drawing a parallel to the well-known Lindblad master equation (ME). In \cref{sec:method} we detail the proposed characterisation scheme, by recalling the exact formula for the correlation functions, how to compute them efficiently numerically, and explaining the fitting procedure. In \cref{sec:examples} we demonstrate the approach in numerical simulations by estimating the parameters of several quantum systems: a driven anharmonic oscillator measured by heterodyne detection, a driven two-level system under photodetection, and a recent experiment continuously monitoring a two-photon dissipative oscillator \cite{berdou2023one}. \Cref{sec:discussion} compares the method with existing approaches and discusses its advantages and limitations. Finally, \cref{sec:conclusion} presents conclusions.

\section{The SME formalism}\label{sec:sme}

An observer describes the state of an open quantum system by a mixed quantum state, the density matrix $\rho_t$ at time $t$. Mathematically, we associate with the system a Hilbert space $\Hcal$ of dimension $n_\Hcal$ (possibly infinite), and the quantum state is a Hermitian positive semidefinite operator with unit trace: ${\rho\in\Omega_\Hcal=\{\rho\in\Hcal,\ \rho\geq0, \Tr{\rho}=1\}}$. Different laws of evolution describe the trajectory of the quantum state $\{\rho_t\}_{t\in[0,T]}$ in state space $\Omega_\Hcal$. In the Markovian scenario, there are two situations: if the observer cannot observe the quantum systems composing the environment, the state trajectory is described by the Lindblad master equation (ME). Conversely, if the observer measures these external degrees of freedom, the state trajectory is governed by the stochastic master equation (SME). An effective approach to understanding the link between these two equations is to consider the interactions between the system and its environment in discrete time.

\subsection{The discrete case}

In discrete time, at each time step $k$, the system of interest entangles with another quantum system, which is then measured projectively by a detector. Mathematically, each step is described by a quantum measurement with Kraus decomposition $\{M_r\}$, a set of operators satisfying the normalisation condition ${\sum_r M_r^\dag M_r=\mathrm{I}}$ \cite{nielsen2010quantum}. The index $r$ refers to the different measurement outcomes. We consider three situations: when the measurement outcomes are perfect, unknown or imperfect.

In the first situation, the detector is perfect, and the measurement outcome $r$ is known to the observer. The new state $\rho_{k+1\knowing r}$ is obtained from the previous state $\rho_k$ by applying the corresponding measurement operator $M_r$:
\begin{equation}\label{eq:quantum-measurement}
    \rho_{k+1\knowing r} = \frac{M_r\rho_k M_r^\dag}{\mathrm{Tr}[M_r\rho_k M_r^\dag]}=\frac{\Kcal_r(\rho)}{\mathrm{Tr}[\Kcal_r(\rho)]},
\end{equation}
where the map $\Kcal_r$ is defined by ${\Kcal_r(\bullet)=M_r\bullet M_r^\dag}$. The probability of measuring the outcome $r$ is:
\begin{equation}\label{eq:quantum-measurement-probability}
    \PP[r] = \mathrm{Tr}[\Kcal_r(\rho)].
\end{equation}

In the second situation, the observer does not have access to the detector, and the measurement outcome is unknown. The new state $\rhobar_{k+1}$ is obtained by averaging all possible states $\rho_{k+1\knowing r}$ after measurement, weighted by their probability of occurrence. Using \cref{eq:quantum-measurement,eq:quantum-measurement-probability}:
\begin{equation}\label{eq:cptp}
    \rhobar_{k+1} = \sum_{r} \PP[r]\,\rho_{k+1\knowing r} = \sum_{r} M_r\rho_k M_r^\dag = \overline\Kcal(\rho).
\end{equation}
The resulting evolution is characterised by a completely positive trace-preserving (CPTP) map $\overline\Kcal$ independent of the measurement outcome, defined by ${\overline\Kcal(\bullet)=\sum_{r} M_r\bullet M_r^\dag}$. Averaging over all possible measurement outcomes is empirically equivalent to tracing out the unknown state of the environment. Note that any physically possible evolution, described by a CPTP map, can be reformulated as an unread quantum measurement.

A third situation lies between these two extremes: when the detector is imperfect, yielding a partially informative outcome $p$. Such a detector is characterised by the conditional probability $\PP[p\knowing r]$ of getting the imperfect outcome $p$, given that the true outcome was $r$. The new state $\rhobar_{k+1 \knowing p}$ is obtained by averaging all possible states $\rho_{k+1\knowing r}$ after the measurement, weighted by their probability of occurrence given that the imperfect outcome was $p$:
\begin{equation}\label{eq:imperfect-quantum-measurement}
    \rhobar_{k+1\knowing p} = \sum_{r} \PP[r\knowing p]\,\rho_{k+1\knowing r} = \frac{\overline\Kcal_p(\rho_k)}{\mathrm{Tr}[\overline\Kcal_p(\rho_k)]},
\end{equation}
where the map $\overline\Kcal_p$ for imperfect outcome $p$ is defined by ${\overline\Kcal_p(\bullet) = \sum_{r} \PP[p\knowing r]\,M_r\bullet M_r^\dag}$. This result can be derived using Bayes' theorem and \cref{eq:quantum-measurement,eq:quantum-measurement-probability}. The probability of measuring the imperfect outcome $p$ is ${\PP[p]=\mathrm{Tr}[\overline\Kcal_p(\rho_k)]}$. For perfect measurement outcomes we have $\PP[p\knowing r]=\delta_{pr}$ (where $\delta$ is the Kronecker delta symbol), and we get back \cref{eq:quantum-measurement}. For unknown (completely random) measurement outcomes we have $\PP[p\knowing r]=1/n_p$ (where $n_p$ is the number of imperfect outcomes), and we get back \cref{eq:cptp}.

In the limit of infinitely frequent and infinitely weak projective measurements of the environment, we find a different law for the continuous-time evolution of the state in each situation. In the first situation, with perfect outcomes, the state evolves according to the SME. In the second situation, with unknown outcomes, the state evolves according to the ME. In the third situation, with imperfect outcomes, the state evolves according to the SME, but including detector imperfections such as efficiency and dark count rate, as detailed below.

\subsection{The Lindblad master equation}

The ME describes the evolution of the average state $\rhobar_t$ when the observer does not monitor the environment, for example for a purely dissipative process or for unread measurements. The evolution is deterministic, it is described by the linear ordinary differential equation (ODE):
\begin{equation}\label{eq:lindblad}
    \frac{\dd\rhobar_t}{\dt} = \Lcal_t(\rhobar_t) = -i[H_t,\rhobar_t] + \sum_{k=1}^N\Dcal_{L_{k,t}}(\rhobar_t),
\end{equation}
where $\Lcal_t$ is the Liouvillian superoperator, $H_t$ is the Hamiltonian of the system, $\{L_{1,t},\dots, L_{N,t}\}$ is a collection of jump operators and $\Dcal_L(\bullet) = L\bullet L^\dag - \frac12 L^\dag L\bullet - \frac12 \bullet L^\dag L$ is the standard dissipator. \Cref{eq:lindblad} is the continuous-time formulation of \cref{eq:cptp} in the discrete setting.

For simplicity, we consider in the following a single loss channel with a time-independent jump operator $L$. If this loss channel is fully or partially measured by the observer, the acquired information can be used to update the system state. This leads to a better (less mixed on average) description of the state trajectory than the ME: the stochastic master equation.

\subsection{The stochastic master equation}

The SME describes the evolution of the state $\rho_t$ when the observer continuously measures the loss channel with a detector. The evolution is non-deterministic, it is described by the non-linear stochastic differential equation (SDE):
\begin{equation}\label{eq:sme}
    \dd \rho_t = \Lcal_t(\rho_t)\,\dt + \Mcal_L(\rho_t, \dYt),
\end{equation}
where $\Mcal_L$ is a superoperator which depends on a stochastic process $\dYt$: the measured signal. The first part of the equation is the deterministic Lindblad update. The second part models the stochastic backaction of the measurement, which is continuously taken into account. In the case of a perfect detector, \cref{eq:sme} is the continuous-time formulation of \cref{eq:quantum-measurement} in the discrete setting, and more generally of \cref{eq:imperfect-quantum-measurement} when detector imperfections are taken into account.

The detector output is a continuous-time signal defined by the rate of change of the stochastic process $\dYt$ over time: ${I_t=\dYt/\dt}$. The path followed by the quantum state over time is entirely determined by this measured signal: each time the observer performs a new experiment (labelled $j$), they measure a particular realisation of the stochastic signal $\{I_t^{(j)}\}_{t\in[0, T]}$, which corresponds to a unique trajectory $\{\rho_t^{(j)}\}_{t\in[0, T]}$ in state space. These trajectories are called \emph{quantum trajectories}, and the state $\rho_t$ is said to be \emph{conditioned} on the information measured by the observer up to time $t$.

There are two classes of SMEs: the \emph{jump} SME and the \emph{diffusive} SME. The difference lies in the type of detector: if the output is binary ($0$ or $1$), the evolution is described by the jump SME, and otherwise if the output takes a continuous range of values, the evolution is described by the diffusive SME. In quantum optics, for example, the jump SME models photodetection, while the diffusive SME models homodyne or heterodyne detection. Mathematically, they differ in the stochastic process driving the SME, and in the form of the measurement backaction superoperator $\Mcal_L$.

\textbf{Jump SME} -- Most of the time the detector detects nothing and outputs $0$, and sometimes, upon detection, the detector \emph{clicks}, and outputs $1$. This stochastic process is modelled by the point process $\dYt = \dNt$ with law:
\begin{align}\label{eq:dNt}
    \begin{split}
        \pr{\dNt=0} & = 1 - \pr{\dNt=1},                              \\
        \pr{\dNt=1} & = \left(\theta + \eta\Tr{L\rho_t\Ld}\right)\dt,
    \end{split}
\end{align}
where $\theta\geq0$ is the dark count rate (taking into account false clicks), and $0\leq\eta\leq1$ is the detector efficiency (taking into account missed clicks). The measured signal $I_t=\dNt/\dt$ is the rate of change of the counting process $N_t=\int_0^t \dNt$, which counts the number of jumps occurring in the time interval $[0,t)$. The measurement backaction is defined by \cite{rouchon2022tutorial}:
\begin{align}\label{eq:Mjump}
    \begin{split}
        \Mcal_L(\rho,\dd N) = & ~ \left(\frac{\theta\rho + \eta L\rho\Ld}{\theta + \eta\Tr{L\rho\Ld}} - \rho\right) \\
                                    & \ \ \ \quad\Big(\dd N - \left(\theta + \eta\Tr{L\rho\Ld}\right)\dt\Big).
    \end{split}
\end{align}

\textbf{Diffusive SME} -- The detector output is real-valued, it is described by the Itô process $\dYt$ defined by:
\begin{equation}\label{eq:dYt}
    \dYt = \sqrt{\eta}\Tr{(L+\Ld)\rho_t}\dt + \dWt,
\end{equation}
where $0\leq\eta\leq1$ is the detector efficiency (taking into account measurement imperfections), and $W_t$ is a Wiener process taking independent Gaussian-distributed increments. The measurement backaction is defined by \cite{jacobs2006straightforward}:
\begin{align}\label{eq:Mdiff}
    \begin{split}
        \Mcal_L(\rho,\dd Y) = & ~ \sqrt\eta\Big(L\rho + \rho\Ld - \Tr{(L+\Ld)\rho}\rho\Big) \\
                                    & \quad\Big(\dd Y - \sqrt{\eta}\Tr{(L+\Ld)\rho}\dt\Big).
    \end{split}
\end{align}

In the jump SME case, the quantum trajectory is discontinuous: the state evolves continuously in state space as long as the detector detects nothing, but undergoes a sudden jump upon detection. In the diffusive SME case, the state evolves continuously in state space, following a Brownian motion-like trajectory. The unconditioned trajectory described by the ME is recovered by averaging over all possible realisations of the stochastic process driving the SME (or equivalently, over all possible quantum trajectories or measured signals): $\rhobar_t = \E{\rho_t}$. Here $\mathbb{E}$ is the statistical average over $N_t$ or $W_t$, it is the continuous-time analogue of the average performed in \cref{eq:cptp} in the discrete setting. Note that although the quantum trajectories described by the jump SME and the diffusive SME are very different in nature, the average state trajectory does not depend on the stochastic process averaged over (jump or diffusive).

\section{Fitting correlation functions}\label{sec:method}

This section explains how to estimate the system parameters by fitting correlation functions of the measured signals.

\subsection{Sharp, filtered and binned signal}

We begin by defining three different types of signal: the \emph{sharp} signal $I_t$, the \emph{filtered} signal $I_f$ and the \emph{binned} signal $I_k$. The notation $I_{\scalebox{0.5}{$\blacksquare$}}$ is overloaded for convenience: the nature of the subscript \raisebox{0.2ex}{\scalebox{0.5}{$\blacksquare$}} (real, function or integer) determines the signal type.

\textbf{Sharp signal} -- As explained in the introduction, the signal $I_t$ is a singular quantity. It can be loosely thought of as a series of Dirac delta distributions for the jump SME, and as white noise with a trend for the diffusive SME. We call this continuous-time signal the \emph{sharp} signal.

\textbf{Filtered signal} -- In practice, the measured signal is filtered by an experimental acquisition chain modelled by a linear filter, characterised by its impulse response $f$. The experimentally available quantity is defined by integrating the sharp signal $I_t$ against the function $f$:
\begin{equation}
    I_f = \int f(t) I_t\,\dt = \int f(t)\,\dYt.
\end{equation}
Note that the first integral with $I_t$ is just a formal notation for the second integral with $\dYt$, which is rigorously defined as a point process integral for a jump signal or an Itô integral for a diffusive signal. We call this processed signal the \emph{filtered} signal.

If the acquisition chain is fully analogue, the filtered signal is also continuous in time: it is defined by a continuous set of filter functions indexed by time. In common experimental setups, however, the acquisition chain is completed by a photocounter or an analogue-to-digital converter (ADC), which converts the continuous-time analogue signal into a discrete-time digitised signal $\{I_0, \dots, I_N\}$. Each value $I_k$ is then defined by:
\begin{equation}
    I_k = \int f_k(t) I_t\,\dt,
\end{equation}
where $f_k$ is the $k$-th impulse response of the acquisition chain.

\textbf{Binned signal} -- Digitisation is usually performed by averaging the signal over a duration $\Delta t$ (a time bin), which is longer than the bandwidth of the various acquisition chain components. The filter function for the $k$-th time bin can then be approximated by a rectangular window of duration $\Delta t$.

In the case of the jump SME, the detector typically returns the number of click events over a given time interval. The filter function ${f_k=\1_{[k\Delta t, (k+1)\Delta t)}}$ is the indicator function defined by $\1_\Omega(t)=1$ if $t\in\Omega$ and $\1_\Omega(t)=0$ otherwise. The resulting signal is discrete-valued $I_k\in\mathbb{N}$:
\begin{equation}\label{eq:Ik-jump}
    I_k = \int_{k\Delta t}^{(k+1)\Delta t} I_t\,\dt = N_{[k\Delta t,(k+1)\Delta t)},
\end{equation}
where the counting process $N_{[t_1,t_2)}=\int_{t_1}^{t_2} \dNt$ counts the number of jumps occurring in the time interval $[t_1,t_2)$. In the limit where there is at most one jump per time bin $\Delta t$, the signal is binary $I_k\in\{0,1\}$, it is just a sequence of zeros and ones: $00100010\dots$. The filtering takes into account the (inevitably finite) time resolution of the detector.

In the case of the diffusive SME, the signal is typically averaged over some duration $\Delta t$. The filter function is ${f_k = G/\Delta t\ \1_{[k\Delta t, (k+1)\Delta t)}}$, where $G$ is the gain of the acquisition chain. The resulting signal is real-valued $I_k\in\mathbb{R}$:
\begin{equation}\label{eq:Ik-diff}
    I_k = \frac{G}{\Delta t} \int_{k\Delta t}^{(k+1)\Delta t} I_t\,\dt.
\end{equation}

\begin{figure}[htb]
    \centering
    \includegraphics{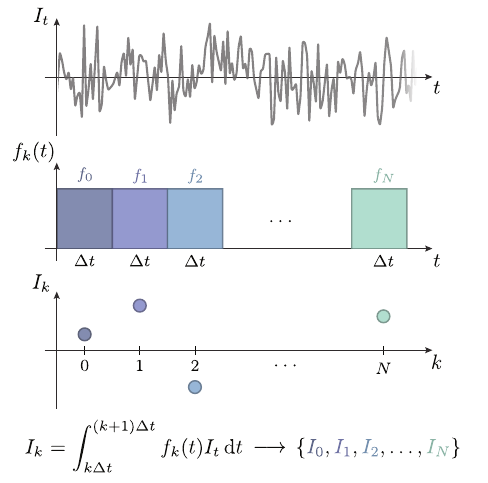}
    \caption{\textbf{Acquisition chain digitisation.} Filtering and digitising the sharp diffusive signal $I_t$ against the rectangular window filter functions $f_k$ results in a discrete-time binned signal $\{I_0,\dots,I_N\}$.}
    \label{fig:binning}
\end{figure}

In both cases, we call this averaged signal the \emph{binned} signal. From a practical point of view, this discrete-time signal $I_k$ is the only quantity available to the experimentalist. This digitisation process is illustrated in \Cref{fig:binning} for the diffusive SME.

\subsection{Correlation functions}

A random variable $X$ is fully characterised by its moments $\E{X^n}$. Similarly, a stochastic process is fully characterised by its correlation functions. The $n$-point correlation function of the sharp signal $I_t$ at times ${t_1<\dots<t_n}$ is defined by $\E{I_{t_1}\dots I_{t_n}}$, where $\mathbb{E}$ is the statistical average over the stochastic process driving the SME (as defined at the end of \cref{sec:sme}).

A naive way to compute the correlation functions would be to sample a large number of trajectories, and explicitly calculate $\E{I_{t_1}\dots I_{t_n}}$ by averaging over all realisations. This approach is prohibitively expensive numerically, and inaccurate due to statistical errors. Fortunately, there is an exact formula to compute any correlation function directly from the SME, without sampling any trajectory. It has been derived for restricted cases in several recent and older works \cite{barchielli1991measurements,korotkov2001output,wiseman2009quantum,xu2015correlation,diosi2016structural,jordan2016anatomy,foroozani2016correlations,atalaya2018multitime}. In 2018, a general derivation for the diffusive SME was independently discovered by \cite{tilloy2018exact} and \cite{hagele2018higher}. Related calculations can also be found in the field of condensed matter physics, which focuses on full counting statistics, cumulants and spectral representation \cite{flindt2010counting,landi2024current}.

However, as we explained previously, the only quantity actually available to the experimentalist is the filtered or binned signal $I_f$, and therefore one can only estimate the correlation functions of this signal $\E{I_{f_1} I_{f_2} \dots I_{f_n}}$ experimentally. The derivation of an exact formula for these quantities was proposed in \cite{tilloy2018exact} for the diffusive SME. It was extended for both types of SMEs in 2023, with a new self-contained and more accessible derivation, together with an efficient numerical method to compute these quantities \cite{guilmin2023correlation}.

This subsection provides a detailed guide to the calculation of correlation functions, divided into three parts. In the first part, \hyperref[sec:theory]{\ref*{sec:theory} Theory}, we recall the formulae for the correlation functions of the sharp signal (see \cref{eq:sharp-correlation}) and the filtered and binned signals (see e.g. \cref{eq:f2-correlation}). The second part, \hyperref[sec:numerics]{\ref*{sec:numerics} Numerics}, discusses how to efficiently compute these quantities numerically. The third part, \hyperref[sec:experimental]{\ref*{sec:experimental} Experimental}, describes how to estimate them experimentally. For simplicity, the presentation focuses primarily on a single signal, with a time-independent jump operator $L$. However, we also explain the straightforward generalisation to compute correlation functions between signals measured by different detectors.

\subsubsection{Theory}\label{sec:theory}

We first introduce the \emph{correlation superoperator} $\Ccal_L$, which is ubiquitous in the correlation functions formulae. Its precise form depends on the detector type:
\begin{align}
    \Ccal_L(\bullet) & = \theta \bullet +\,\eta L\bullet L^\dag &  & \text{for the jump SME},      \\
    \Ccal_L(\bullet) & = \sqrt\eta (L\bullet + \bullet L^\dag)  &  & \text{for the diffusive SME}.
\end{align}

\textbf{Generating functional} -- The moments of a random variable can be obtained from the \emph{moment-generating function} ${M_t=\E{e^{tX}}}$, whose $n$-th derivative gives the $n$-th moment of $X$:
\begin{equation}
    \E{X^n} = \frac{\dd^nM}{\dt^n}\Big\vert_{t=0}.
\end{equation}
Similarly, the correlation functions of a stochastic process can be obtained by taking derivatives of the \emph{generating functional} $\Zcal(j)$ with $j$ a function \cite{barchielli2009quantum, tilloy2018exact}:
\begin{equation}
    \Zcal(j) = \E{\exp\left(\int_0^{\infty} j_tI_t\,\dt\right)}.
\end{equation}
The main result of \cite{tilloy2018exact, guilmin2023correlation} is that $\Zcal(j)$ can be expressed directly from the SME and the initial state, with the explicit formula:
\begin{equation}\label{eq:Zcal}
    \Zcal(j) = \Tr{\mathscr{T}\exp\left(\int_0^\infty \Lcal^j_t\,\dt\right) (\rho_0)},
\end{equation}
where $\mathscr{T}$ is the time-ordering symbol, and $\Lcal^j_t$ is a superoperator whose precise form depends on the detector type:
\begin{align}
    \Lcal^j_t & = \Lcal_t + (e^{j_t}-1)\,\Ccal_L              &  & \text{for the jump SME},\label{eq:Lcalj-jump}      \\
    \Lcal^j_t & = \Lcal_t + j_t\,\Ccal_L+\frac{j_t^2}{2} &  & \text{for the diffusive SME}\label{eq:Lcalj-diff}.
\end{align}
We propose yet another derivation of \cref{eq:Zcal} in \cref{app:new-proof}, -- the shortest to date --, by directly applying Itô's lemma to the SME.

\textbf{Sharp signal} -- The sharp signal correlation functions are obtained by taking functional derivatives of the generating functional $\Zcal(j)$ with respect to $j_{t_1},\dots,j_{t_n}$:
\begin{equation}
    \E{I_{t_1}\dots I_{t_n}} = \frac{\delta^n\Zcal}{\delta j_{t_1}\dots\delta j_{t_n}}\Big\vert_{j=0}.
\end{equation}
For example, using \cref{eq:Zcal}, the explicit formula for different times ${t_1<\dots<t_n}$ and for a constant Liouvillian $\Lcal_t=\Lcal$ is given by:
\begin{equation}\label{eq:sharp-correlation}
    \E{I_{t_1}\dots I_{t_n}} = \Tr{\Ccal_L e^{(t_n-t_{n-1})\Lcal} \dots \Ccal_L e^{t_1\Lcal} (\rho_0)}.
\end{equation}
The result is the trace of a peculiar operator, which is obtained by evolving the initial state with the usual Lindblad evolution, interspersed with the application of the correlation superoperator at each correlation time $t_k$.

This equation is only valid for different times ${t_1<\dots<t_n}$, because \emph{equal time contributions} yield Dirac delta distribution. These contributions should not be overlooked when evaluating correlation functions of filtered or binned signals.

\textbf{Filtered signal} -- A naive way to compute the $n$-point correlation function of the filtered signal $\E{I_{f_1} \dots I_{f_n}}$ is to explicitly compute the $n$-dimensional integral of $\E{I_{t_1} \dots I_{t_n}}$ against the filter functions:
\begin{align}
    \E{I_{f_1}\dots I_{f_n}}=\int_{t_1}\ccdots\int_{t_n} & \E{I_{t_1} \dots I_{t_n}}\\
    &f_1(t_1)\dots f_n(t_n)\,\dt_1\dots\dt_n\notag,
\end{align}
where $\E{I_{t_1} \dots I_{t_n}}$ is given by \cref{eq:sharp-correlation} when the filter functions $f_k$ have non-overlapping support, and includes additional Dirac delta distributions where they overlap. One could then evaluate this expression numerically by discretising the $n$-dimensional integral and evaluating the integrand at each quadrature point. However, this computation quickly becomes prohibitively expensive for large Hilbert space dimensions, and correlation functions involving more than two points. As detailed in \cite{guilmin2023correlation}, there is a faster and more practical way to compute these quantities by solving modified Lindblad master equations, which we now recall.

The filtered signal correlation functions are obtained by taking partial derivatives of the generating functional $\Zcal(j)$ with $j = \alpha_1 f_1 + \dots + \alpha_n f_n$ with respect to $\alpha_1,\dots,\alpha_n$:
\begin{equation}\label{eq:partial}
    \E{I_{f_1}\dots I_{f_n}} = \frac{\partial^n \Zcal}{\partial\alpha_1\dots\partial\alpha_n} \Big\vert_{\alpha_1,\dots,\alpha_n=0}.
\end{equation}
Let us introduce a time $T$ which is greater than any time in the support of $f_1,\dots,f_n$, such that for all $t\geq T$, we have $j_t=0$. Then, if we identify $\Lcal^j_t$ as the generator of an ordinary differential equation (ODE) in \cref{eq:Zcal}, we can write ${\Zcal(j) = \mathrm{Tr}[\rho_T^j]}$ at the final time $T$, with $\rho_t^j$ solution of:
\begin{equation}\label{eq:ODE}
    \frac{\dd \rho_t^j}{\dt} = \Lcal^j_t(\rho_t^j).
\end{equation}
If we now pull the partial derivatives in \cref{eq:partial} inside the trace, we have:
\begin{align}
    \E{I_{f_1}\dots I_{f_n}} &= \Tr{\frac{\partial^n \rho_T^j}{\partial\alpha_1\dots\partial\alpha_n}\Big\vert_{\alpha_1,\dots,\alpha_n=0}}\\
    &=\Tr{\rho_T^{1,\dots,n}}.
\end{align}
To compute $\rho_T^{1,\dots,n}$, the $n$-th order partial derivative of $\rho_T^j$ with respect to $\alpha_1,\dots,\alpha_n$, we (manually) forward differentiate the ODE \cref{eq:ODE}:
\begin{equation}
    \frac{\dd \rho_t^{1,\dots,n}}{\dt} = \frac{\partial^n \left(\Lcal^j_t(\rho_t^j)\right)}{\partial\alpha_1\dots\partial\alpha_n}\Big\vert_{\alpha_1,\dots,\alpha_n=0}.
\end{equation}
Recursively distributing the derivatives results in an augmented system of $2^n$ coupled linear ODEs (called sensitivity equations), which includes all unordered combinations of partial derivatives with respect to subsets of $\{\alpha_1,\dots,\alpha_n\}$. Each ODE describes the evolution of a fictitious state under the regular Lindblad evolution, with the introduction of additional coupling \emph{source terms}. The system is solved from time $0$ to time $T$.

For example, the two-point correlation function ${\E{I_{f_1}I_{f_2}}}$ is found by evolving four fictitious states $(\rho_t, \rho_t^1, \rho_t^{2}, \rho_t^{12})$ with initial condition $(\rho_0, 0, 0, 0)$. The result is given by the trace of $\rho^{12}_T$ at the final time $T$: ${\E{I_{f_1}I_{f_2}}=\Tr{\rho^{12}_T}}$. The system of coupled linear ODEs describing the evolution of this large state can be summarised by its generator in matrix form. For example in the diffusive case (dropping the time index for compactness):
\begin{equation}\label{eq:f2-correlation}
    \begin{bmatrix}\dot\rho \\ \dot\rho^1 \\ \dot\rho^2 \\ \dot\rho^{12}\end{bmatrix}=
    \begin{bmatrix}
        \Lcal        & 0            & 0            & 0     \\
        f_1\,\Ccal_L & \Lcal        & 0            & 0     \\
        f_2\,\Ccal_L & 0            & \Lcal        & 0     \\
        f_1f_2       & f_2\,\Ccal_L & f_1\,\Ccal_L & \Lcal
    \end{bmatrix}
    \begin{bmatrix}\rho \\ \rho^1 \\ \rho^2 \\ \rho^{12}\end{bmatrix}.
\end{equation}
For convenience, we give in \cref{app:filtered} the system of ODEs to solve for correlation functions up to order three, for both types of SME.

\textbf{Binned signal} -- The correlation functions of the binned signal $I_k$ are just a special case where the filter functions are rectangular windows ${f_k\propto\1_{[k\Delta_t, (k+1)\Delta_t)}}$. This can be numerically advantageous, as explained below.

\textbf{Correlation functions between different signals} -- These formulae can be easily generalised to correlation functions between signals from different detectors, which can be of jump and/or diffusive type. The generating functional has exactly the same expression, where $j$ is now a set of functions, one for each detector: $j=\{j^\mu\}_{1\leq\mu\leq n_\mu}$ where $n_\mu$ is the number of measured signals. Splitting the detectors into those of jump-type $\mu\in S_\text{jump}$ and those of diffusive-type $\mu\in S_\text{diffusive}$, the new generator of the evolution in \cref{eq:Zcal} is:
\begin{align}
    \begin{split}
        \Lcal^j_t = \Lcal_t & + \sum_{\mu\in S_\text{jump}} (e^{j^\mu_t}-1)\,\Ccal_{L^\mu}                                       \\
                            & + \sum_{\mu\in S_\text{diffusive}} \left(j^\mu_t\,\Ccal_{L^\mu}+\frac{(j^\mu_t)^2}{2}\right).
    \end{split}
\end{align}

For example, the sharp signal formula \cref{eq:sharp-correlation} extends directly to signals from different detectors, $I_{t_1}^{\mu_1}, \dots, I_{t_n}^{\mu_n}$, by inserting the corresponding correlation superoperator $\Ccal_{L^{\mu_k}}$ at each time $t_k$. Similarly, the filtered signal formula \cref{eq:f2-correlation} extends to two signals from different diffusive detectors, $I_{f_1}^{\mu_1}$ and $I_{f_2}^{\mu_2}$:
\begin{equation}
    \begin{bmatrix}\dot\rho \\ \dot\rho^1 \\ \dot\rho^2 \\ \dot\rho^{12}\end{bmatrix}=
    \begin{bmatrix}
        \Lcal        & 0            & 0            & 0     \\
        f_1\,\Ccal_{L^{\mu_1}} & \Lcal        & 0            & 0     \\
        f_2\,\Ccal_{L^{\mu_2}} & 0            & \Lcal        & 0     \\
        \delta_{\mu_1\mu_2}f_1f_2       & f_2\,\Ccal_{L^{\mu_2}} & f_1\,\Ccal_{L^{\mu_1}} & \Lcal
    \end{bmatrix}
    \begin{bmatrix}\rho \\ \rho^1 \\ \rho^2 \\ \rho^{12}\end{bmatrix},
\end{equation}
where $\delta$ is the Kronecker delta symbol.

\subsubsection{Numerics}\label{sec:numerics}

\textbf{Sharp signal} -- The sharp signal correlation functions \cref{eq:sharp-correlation} can be computed by evolving the initial state with the ME using any regular solver, and applying the correlation superoperator at each correlation time. The numerical cost is the same as solving a single ME, up to the final correlation time $t_n$.

\textbf{Filtered signal} -- The system of ODEs can be viewed as a single large linear ODE, which is integrated using commonly available ODE solvers, such as high-order Runge-Kutta schemes. Note that for large Hilbert space dimensions, it is not recommended to explicitly store or diagonalise the full matrix as written in \cref{eq:f2-correlation}, both for memory and runtime reasons. A better solution is to apply the generator to the large state at each time step, which only costs $\mathcal{O}(n_\Hcal^2)$ in memory and $\mathcal{O}(n_\Hcal^3)$ in time (where $n_\Hcal$ is the Hilbert space dimension). For low-order correlation functions, the numerical cost is the same as solving a few MEs up to time $T$.

\textbf{Binned signal} -- If the filter functions are rectangular windows and the Liouvillian is time-independent, the generator of the large ODE is piecewise constant in time. It can be integrated by successive exponentiation over each time interval. For large Hilbert space dimensions, it is not necessary to compute the exponential explicitly, but simply its action on an operator. This can be done efficiently using e.g. Krylov subspace methods. For low-order correlation functions, the numerical cost is the same as solving a few MEs with a time-independent Liouvillian up to time $T$.

\textbf{Back to the sharp signal} -- If the digitisation time $\Delta t$ is very small compared to the fastest timescale of the system, the correlation functions of the binned signal can be \emph{approximated} with the sharp signal formula \cref{eq:sharp-correlation}. For example for the binned diffusive signal $I_k$ defined in \cref{eq:Ik-diff}, the approximation is ${\E{I_{k_1}\dots I_{k_n}}\approx G^n\,\mathbb{E}[I_{t_{k_1}'}\dots I_{t_{k_n}'}]}$, where $I_{t_k'}$ is the sharp signal at the center of the time bin $t_k'=k\Delta t+\Delta t/2$. Importantly, this calculation is not valid for coincident bins. Moreover, particular care must be taken to ensure that $\Delta t$ is sufficiently small for the approximation to be correct. A good way to verify the validity of the approximation is to compute the exact formula and check that it matches.

\Cref{app:numerical} discusses additional numerical considerations: (i) how to exploit the block structure of the ODE generator to reduce the numerical cost to that of solving a single ME, (ii) how to efficiently vectorise the computation of multiple correlation functions at different times, and (iii) how to appropriately normalise the generator when using adaptive step-size ODE solvers.

\subsubsection{Experimental}\label{sec:experimental}

\begin{figure}[htb]
    \centering
    \includegraphics{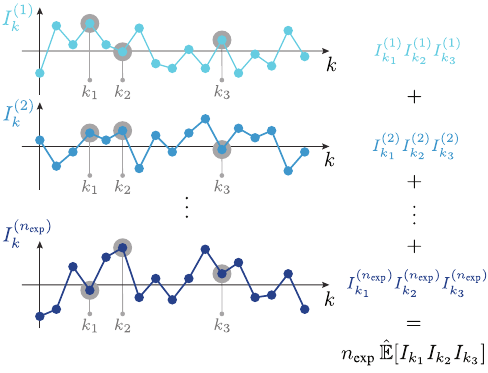}
    \caption{\textbf{Experimental estimate of correlation functions.} The three-point correlation function $\E{I_{k_1}I_{k_2}I_{k_3}}$ of a digitised diffusive signal is estimated experimentally by averaging over multiple realisations of the experiment.}
    \label{fig:cf}
\end{figure}

Estimating the correlation functions of the measured signal is straightforward experimentally. For each realisation of the experiment (labelled $j$), the experimentalist has access to a new digitised signal $\{I_0^{(j)},\dots,I_N^{(j)}\}$. The $n$-point correlation function is given by:
\begin{equation}
    \E{I_{k_1}\dots I_{k_n}} = \lim_{\nexp\to\infty}\frac{1}{\nexp}\sum_{j=1}^\nexp I_{k_1}^{(j)} \dots I_{k_n}^{(j)},
\end{equation}
where $\mathbb{E}$ is the statistical average over an infinite number of realisations $\nexp$ of the experiment (as defined at the end of \cref{sec:sme}). In practice, the correlation functions are estimated with some statistical error due to the finite number of realisations. We denote $\Ehat{I_{k_1}\dots I_{k_n}}$ the experimental estimate of the expectation value:
\begin{equation}\label{eq:Ehat}
    \Ehat{I_{k_1}\dots I_{k_n}} = \frac{1}{\nexp}\sum_{j=1}^\nexp I_{k_1}^{(j)}  \dots I_{k_n}^{(j)}.
\end{equation}
\Cref{fig:cf} illustrates this calculation for a three-point correlation function.

\Cref{app:experimental} discusses additional experimental considerations: (i) how to estimate the uncertainties of these experimental estimates, (ii) specificities when starting from the system steady state, and (iii) the trade-off involved in choosing the digitisation time $\Delta t$.

\subsection{Least-squares fit}\label{subsec:least-squares}

We assume that the model of the system is known, except for the values of $p$ parameters, gathered in the vector $\btheta_\textbf{true}\in\mathbb{R}^p$. These parameters may appear in different parts of the model, such as the initial state, the Hamiltonian, jump operators, dark count rates or efficiencies, or filter functions. Our goal is to estimate the values of these unknown parameters.

Generally, each parameter has a unique impact on the evolution of the state, and thus on the statistics of the measured signals. To simultaneously estimate the $p$ parameters, first some correlation functions of interest are experimentally estimated, and then the exact formula is used to fit these observations with a least-squares method. The result is a vector of parameters $\btheta_\textbf{fit}$ that minimises the difference between the experimental estimates and the theoretical model. The hope is that the estimated parameters $\btheta_\textbf{fit}$ are close to the true parameters $\btheta_\textbf{true}$. In the next section, we demonstrate the effectiveness of this idea in simulation using three different examples. We now discuss potential challenges and methodological considerations.

\textbf{Degeneracy and local minima} -- A number of situations can lead to an incorrect estimation of the parameters. First, the parameters we are trying to estimate may be degenerate: different combinations of parameters give the same correlation functions. We discuss different types of degeneracy -- fundamental non-identifiability, low-order degeneracy, and statistical degeneracy --, how to identify them, and possible workarounds in \cref{app:degeneracy}. The third example (\cref{sec:ex3}) illustrates a real-life parameter degeneracy, and how to work around it. Second, if the problem is not convex and the initial parameters guess is far from the true values, the minimisation may get stuck in a local minima. If the parameters are not degenerate -- which is the first thing to check --, the goodness of the fit will be visually poor. One solution is to run the fitting procedure several times with different initial guesses (which can be done numerically in parallel), and keep the best-fitting parameters.

\textbf{Error bars} -- After the fit has converged, there remains uncertainty in the parameter values due to the approximate estimation of the correlation functions from the experimental data. A practical way to compute error bars for each parameter is to use subsampling: the $\nexp$ experimental realisations are divided into a number of subsets $\nsubset$, and the fit is run independently for each subset. The standard deviation for each parameter is given by the standard deviation of the $\nsubset$ estimated parameters, scaled by $\sqrt{\nsubset}$. Another option is to use bootstrapping.

\textbf{Correlated residuals} -- The experimental estimates for multiple correlation functions (at different set of times and of different orders) are typically computed from the same pool of realisations and, as a consequence, the fit residuals are correlated. The correct way to perform a nonlinear least-squares fit with correlated data is to solve the \emph{feasible generalised least-squares problem}, which requires estimating the covariance matrix of the residuals to best account for their correlation. For simplicity, this can be ignored initially, and a naive least-square fit can be used. The consequence is that the correlation of the residuals is interpreted as part of the data, resulting in a slightly biased estimate of the parameters. Note also that the error bars returned by the fit algorithm will be underestimated, and one should rather use the subsampling procedure suggested above to obtain valid uncertainties on the estimated parameters.

\textbf{Numerics} -- For the examples in this paper, we solved the minimisation problem with the Levenberg-Marquardt algorithm using the \texttt{least\_squares()} function from the SciPy library \cite{virtanen2020scipy}. For simplicity, we did not account for the correlation of the residuals. The fit converges in less than a minute on a standard CPU for typical bosonic problems.

Access to the gradient can significantly improve both the fitting time and accuracy when estimating many parameters simultaneously. It can also help in designing experiments that are sensitive to all parameters (see \cref{app:degeneracy}). The gradient can be computed numerically using modern automatic differentiation libraries to integrate the system of ODEs. For the examples discussed in this paper, we used the \texttt{Tsit5} ODE solver from the Diffrax library \cite{kidger2022on} (based on JAX \cite{jax2018github}) to compute the correlation functions. However, we did not use the gradient capabilities, because it was not necessary for the selected examples, and to facilitate easy replication of our work with any standard library of ODE solvers.

\section{Numerical examples}\label{sec:examples}

We demonstrate the method using three examples, each with its own peculiarities. In the first example (\cref{sec:ex1}), we characterise a driven anharmonic oscillator under heterodyne detection. The second example (\cref{sec:ex2}) illustrates the method in the jump case: we fit the statistics of a photocounting signal to estimate the parameters of a driven two-level system. We use this example to explain how to develop an intuition about how each parameter is estimated from the measured data. The last example (\cref{sec:ex3}) is inspired by a recent experiment \cite{berdou2023one}, in which a two-photon dissipative oscillator implemented on a superconducting circuit platform was continuously monitored by homodyne detection. This example illustrates an interesting real-life parameter degeneracy, and ways to overcome it.

The examples are based on synthetic data generated by numerical simulation. For each system, we simulate the SME with a small time step $\delta t$ for a large number of quantum trajectories. We then integrate the simulated signals over a duration $\Delta t\gg\delta t$ to obtain discrete-time binned signals as defined by \cref{eq:Ik-jump,eq:Ik-diff}. Finally, we estimate the correlation functions of these binned signals by averaging over all the simulated trajectories. These estimates are representative of what would be obtained in a real experiment.

The stochastic trajectories are simulated using a first-order weak convergence numerical method that preserves the CPTP nature of the evolution up to machine precision \cite{rouchon2014models, rouchon2015efficient, jordan2016anatomy, rouchon2022tutorial}. For efficiency, all the trajectories are simulated simultaneously by stacking the states in a single large array. We run the simulation on a GPU (an NVIDIA L40S card) using the new open-source Python library \href{https://github.com/dynamiqs/dynamiqs}{Dynamiqs} \cite{guilmin2024dynamiqs}, designed for high-performance quantum systems simulation with JAX \cite{jax2018github}. Note that access to a GPU is not required to fit the data from a real experiment, as there is no need to simulate any trajectory, and the fit easily runs on a regular CPU.

\subsection{Heterodyne measurement of a driven anharmonic oscillator}\label{sec:ex1}

The first system we consider is a driven quantum anharmonic oscillator whose fluorescence is monitored by heterodyne detection. We show that the system parameters can be estimated by fitting the two-point correlation functions of the binned heterodyne signals.

\textbf{System} -- The Hilbert space associated with the system is the infinite-dimensional space spanned by the Fock states ${\Hcal=\text{span}\,\{\ket0,\ket1,\dots\}}$ where $\ket{n}$ is the $n$-th photon Fock state. The oscillator bosonic annihilation operator is denoted $a$. The oscillator is driven at resonance by a constant linear drive with complex amplitude $\epsilon=\epsilon_x+i\epsilon_y$, and its anharmonicity is modelled by a self-Kerr nonlinearity $K$. It loses photons with a single-photon loss rate $\kappa$. The system dynamics are described by the following Hamiltonian and jump operator:
\begin{align}
    \begin{split}
        H & = - K/2\,a^{\dag2}a^2 + \epsilon^* a + \epsilon a^\dag, \\
        L & = \sqrt{\kappa}\,a.
    \end{split}
\end{align}
The loss channel $L$ is monitored with total efficiency $\eta$ by heterodyne detection along the quadratures ${X=(a+a^\dag)/2}$ and ${P=i(a^\dag-a)/2}$, see the diffusive SME \cref{eq:sme,eq:Mdiff}. This results in two binned signals $\{I_0^X,\dots,I_N^X\}$ and $\{I_0^P,\dots,I_N^P\}$ for each realisation of the experiment. The heterodyne measurement is modelled by splitting the loss operator $L$ into two parts with halved rates ${L^X=\sqrt{\kappa/2}\,a}$ and ${L^P=\sqrt{\kappa/2}\,(-ia)}$, whose diffusive measurements give the binned signals $I_k^X$ and $I_k^P$ as defined by \cref{eq:dYt,eq:Ik-diff}:
\begin{align}
    \begin{split}
        \dYt^X & = \sqrt{\eta\kappa/2}\,\Tr{(a+a^\dag)\rho_t}\dt + \dWt^X,      \\
        \dYt^P & = \sqrt{\eta\kappa/2}\,\Tr{i(a^\dag-a)\rho_t}\dt + \dWt^P,     \\
        I_k^X  & = \frac{G}{\Delta t}\int_{k\Delta t}^{(k+1)\Delta t} \dYt^X, \\
        I_k^P  & = \frac{G}{\Delta t}\int_{k\Delta t}^{(k+1)\Delta t} \dYt^P,
    \end{split}
\end{align}
where $\dWt^X$ and $\dWt^P$ are independent Wiener processes. We assume that the system has reached its steady state $\rho_\infty$, which is also implicitly fitted since it depends on the parameters.

\textbf{Parameters to be estimated} -- There are six parameters to be estimated experimentally: $K$, $\epsilon_x$, $\epsilon_y$, $\kappa$, $G$ and $\eta$. We assume that the decay rate $\kappa$ has already been determined by a basic spectroscopic experiment. The gain is estimated by turning off the drive and fitting the autocorrelation of the measured signal when the system remains in vacuum (see \cref{app:gain} for more details). We set $G=1.0$ for convenience. We encourage readers with experimental expertise to pause for a moment and consider the sequence of experiments they would design to estimate the values of the four remaining parameters $\btheta=(K, \epsilon_x, \epsilon_y, \eta)$.

\textbf{Results} -- To model experimental data, we simulate ${\nexp=10^6}$ stochastic trajectories, and for each trajectory (labelled $j$), we average the measured signals over $21$ time bins of duration $\Delta t=1/(6\kappa)$, to obtain two binned signals: $\{I_0^{X,(j)},\dots,I_{20}^{X,(j)}\}$ and $\{I_0^{P,(j)},\dots,I_{20}^{P,(j)}\}$. These signals are representative of experimental data, they are the only quantities available to an experimentalist.

\begin{figure}[htb]
    \centering
    \includegraphics{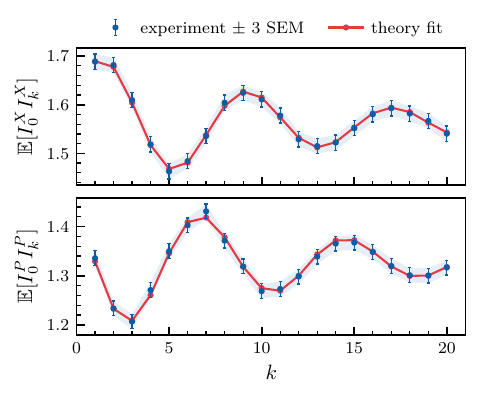}
    \vspace{-0.4cm}
    \caption{\textbf{Joint fit of the heterodyne signals two-point correlation functions.} Experimental estimates (synthetic data, blue points) of the two-point correlation functions of the signal $I^X_k$ (top panel) and the signal $I^P_k$ (bottom panel) are simultaneously fitted with a theoretical model (red line). The blue error bars indicate a range of plus or minus three times the standard error of the mean (SEM) of the experimental estimates.}
    \label{fig:ex1/fit}
\end{figure}

We estimate the two-point correlation functions of the signals at unequal time bins $\E{I_0I_k}\ (1\leq k\leq 20)$ by averaging over the $10^6$ trajectories. We then jointly fit the estimated correlation functions $\Ehat{I_0^XI_k^X}$ and $\Ehat{I_0^PI_k^P}$ with a least-squares method. The four parameters ${\btheta=(K, \epsilon_x, \epsilon_y, \eta)}$ can be estimated simultaneously from these data. The result of the fit is shown in \cref{fig:ex1/fit} and the estimated parameter values are summarised in \cref{table:ex1}. The errors in the estimated values are due to the statistical uncertainty in the estimates $\Ehat{I_0^XI_k^X}$ and $\Ehat{I_0^PI_k^P}$, and would be reduced by averaging over more trajectories. \Cref{app:ex1} gives more details on the system, the simulated data and the fitting procedure.

\begin{table}[htb]
    \centering
    \begin{tabular}{|c|l|l|l|}
        \hline
                        & \textbf{Parameter} & \textbf{Value}      & \textbf{Estimated}\\\hline\hline
        $K$          & self-Kerr          & $100\ \mathrm{kHz}$ & $100\ \ \pm\ 1\ \mathrm{kHz}$ \\
        $\epsilon_x$ & drive (real)       & $300\ \mathrm{kHz}$ & $299\ \ \pm\ 2\ \mathrm{kHz}$\\
        $\epsilon_y$ & drive (imaginary)  & $400\ \mathrm{kHz}$ & $398\ \ \pm\ 2\ \mathrm{kHz}$\\
        $\kappa$     & photon loss rate   & $100\ \mathrm{kHz}$ & \hspace{2.4em}--                    \\
        $\eta$       & efficiency         & $0.8$               & $0.80\ \pm\ 0.01$              \\\hline
    \end{tabular}
    \caption{\textbf{Estimated parameters.} Result of the least-squares fit to simultaneously estimate the four parameters ${\btheta=(K, \epsilon_x, \epsilon_y, \eta)}$. The values are given with a factor of $1/(2\pi)$ where applicable. The standard deviation for each parameter is estimated by subsampling over $\nsubset=10$ subsets.}
    \label{table:ex1}
\end{table}

Fitting correlation functions of the measured signals thus provides a practical way of characterising a monitored driven anharmonic oscillator.

\subsection{Jump measurement of a driven two-level system} \label{sec:ex2}

The next system we study is a driven two-level system (a qubit) whose loss channel is monitored by a photodetector. We show that the system parameters can be estimated by fitting the one-point correlation functions of the binned jump signal. We also explain how to develop an intuition about how each parameter is fitted from the data.

\textbf{System} -- The Hilbert space associated with the system is $\Hcal=\text{span}\, \{\ket0, \ket1\}$ where $\ket0$ is the ground state, and $\ket1$ is the excited state. We describe the dynamics with the Pauli matrices $\sigma_x$, $\sigma_z$ and $\sigma_-$. The system is driven by a real drive with amplitude $\Omega$, which is detuned from the qubit frequency by an amount $\Delta$. The system dynamics are described by the following Hamiltonian and jump operator:
\begin{align}
    \begin{split}
        H & = \Delta \sigma_z + \Omega \sigma_x, \\
        L & = \sqrt{\gamma}\,\sigma_-.
    \end{split}
\end{align}
The loss channel $L$ is monitored by a photodetector with dark count rate $\theta$ and efficiency $\eta$, see the jump SME \cref{eq:sme,eq:Mjump}. This results in a binned signal $\{I_0,\dots,I_N\}$ for each realisation of the experiment. This signal is defined by \cref{eq:dNt,eq:Ik-jump}:
\begin{equation}
    I_k = \int_{k\Delta t}^{(k+1)\Delta t} \dNt.
\end{equation}
We consider the system starting from the excited state ${\rho_0=\ket1\bra1}$.

\textbf{Parameters to be estimated} -- There are five parameters to be estimated experimentally: $\btheta=(\Delta, \Omega, \gamma, \theta, \eta)$. Typically, a specific series of measurements would be designed to characterise each parameter in turn. We show that fitting the mean of the measured signal at different times is an alternative practical method to estimate all the parameters at once.

\begin{figure}[htb]
    \centering
    \includegraphics{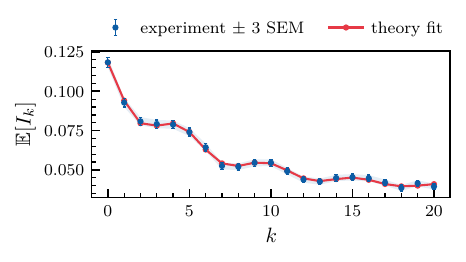}
    \vspace{-0.4cm}
    \caption{\textbf{Fit of the photodetection one-point correlation functions.} Experimental estimates (synthetic data, blue points) of the one-point correlation functions of the signal $I_k$ are fitted by a theoretical model (red line). The blue error bars indicate a range of plus or minus three times the SEM of the experimental estimates.}
    \label{fig:ex2/fit}
\end{figure}

\textbf{Results} -- To model experimental data, we simulate ${\nexp=10^5}$ stochastic trajectories, and for each trajectory (labelled $j$), we average the measured signals over $21$ time bins of duration $\Delta t\approx16\ \mu\text{s}$ (which corresponds to ${1/(2\Delta)}$) to obtain the binned signal $\{I_0^{(j)},\dots,I_{20}^{(j)}\}$. We estimate the mean signal value for each time bin $\E{I_k}\ (0\leq k\leq 20)$ by averaging over the $10^5$ trajectories. We then fit the estimated mean $\Ehat{I_k}$ with a least-squares method, which allows to estimate the five parameters $\btheta=(\Delta, \Omega, \gamma, \theta, \eta)$ simultaneously. The result of the fit is shown in \cref{fig:ex2/fit} and the estimated parameter values are summarised in \cref{table:ex2}. \Cref{app:ex2} gives more details about the system, the simulated data and the fitting procedure.

\begin{table}[htb]
    \centering
    \begin{tabular}{|c|l|l|l|}
        \hline
                        & \textbf{Parameter} & \textbf{Value}      & \textbf{Estimated}\\\hline\hline
        $\Delta$          & detuning          & $5\ \mathrm{kHz}$ & $5.04\ \pm\ 0.03\ \mathrm{kHz}$  \\
        $\Omega$ & drive       & $3\ \mathrm{kHz}$ & $2.99\ \pm\ 0.05\ \mathrm{kHz}$ \\
        $\gamma$     & loss rate   & $2\ \mathrm{kHz}$ & $2.03\ \pm\ 0.06\ \mathrm{kHz}$                    \\
        $\theta$       & dark count rate         & $300\ \mathrm{Hz}$               & $307\ \ \pm\ 7\ \mathrm{Hz}$               \\
        $\eta$       & efficiency         & $0.5$               & $0.49\ \pm\ 0.01$ \\\hline
    \end{tabular}
    \caption{\textbf{Estimated parameters.} Result of the least-squares fit to estimate the five parameters ${\btheta=(\Delta, \Omega, \gamma, \theta, \eta)}$ simultaneously. The values are given with a factor of $1/(2\pi)$ where applicable. The standard deviation for each parameter is estimated by subsampling over $\nsubset=10$ subsets.}
    \label{table:ex2}
\end{table}

\begin{figure}[htb]
    \centering
    \includegraphics{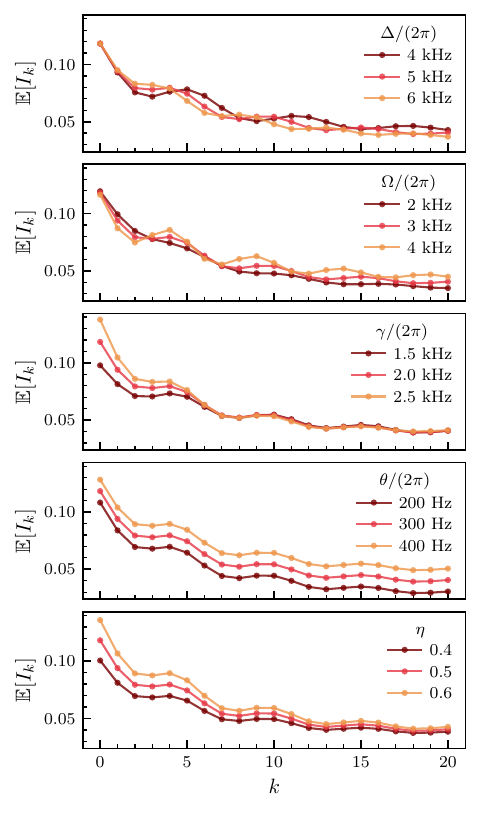}
    \vspace{-0.4cm}
    \caption{\textbf{Correlation function dependence on each parameter.} Each panel shows how the fitted correlation functions changes when one of the parameters is varied (from top to bottom: $\Delta$, $\Omega$, $\gamma$, $\theta$, $\eta$) while all the others are held constant. The middle red curve is the main text fit value.}
    \label{fig:ex2/vary}
\end{figure}

\textbf{Interpreting the method} -- To understand how each parameter is estimated from the measured data, a simple approach is to plot how the fitted correlation functions vary when the value of one of the parameters is changed while all the others are held constant, as illustrated in \cref{fig:ex2/vary}. This allows to visually identify the distinct contributions of each parameter to the fitted data: the detuning $\Delta$ changes the oscillations period, the drive $\Omega$ is responsible for their amplitude, the decay rate $\gamma$ affects the initial value and subsequent exponential decay, the dark count rate $\theta$ causes an overall shift, and the efficiency $\eta$ has yet another unique influence. Having an interpretable characterisation method is very useful for debugging the process, building confidence in the estimated parameters, and possibly adjusting the model of the system if the theory does not fit the measured data.

Fitting the statistics of the photocounting signal is a simple way to fully characterise this two-level system.

\subsection{Homodyne measurement of a two-photon dissipative oscillator}\label{sec:ex3}

We now turn to a recent experiment that proved particularly challenging to characterise \cite{berdou2023one}. For this example, we do not attempt to estimate all the system parameters, but rather show how our method could have been used as the final step in the characterisation process.

This example is interesting for several reasons: the system Hilbert space is of relatively high dimension, the measurement efficiency is particularly low, and there is a low-order degeneracy: the one-, two- and three-point correlation functions are insufficient to estimate the system parameters. We propose two different solutions to lift this degeneracy and to complete the characterisation of the system.

\textbf{System} -- The goal of the experiment is to demonstrate the stabilisation of opposite phase coherent states in the cavity field of a superconducting resonator. We focus on the reduced model of the system, which describes only the dynamics of the cavity. The Hilbert space $\Hcal$ associated with the system is the infinite-dimensional space spanned by the Fock basis as in the first example (\cref{sec:ex1}), and we denote $a$ the annihilation operator of the cavity field mode. The stabilisation is achieved by engineering a peculiar dissipation between the system and its environment, modelled by the jump operator ${L_2=\sqrt{\kappa_2}\,(a^2-\alpha_2^2)}$ with two-photon dissipation rate $\kappa_2$ and complex amplitude $\alpha_2$ \cite{mirrahimi2014dynamically}. The cavity also loses single photons at a rate $\kappa_1$. The system dynamics are described by the following Hamiltonian and jump operators:
\begin{equation}\label{eq:HL1L2}
    \begin{split}
        H   & = 0,                               \\
        L_1 & = \sqrt{\kappa_1}\,a,                \\
        L_2 & = \sqrt{\kappa_2}\,(a^2-\alpha_2^2).
    \end{split}
\end{equation}
The loss channel $L_1$ is monitored with total efficiency $\eta$ by heterodyne detection along the $X$ and $P$ quadratures, see the diffusive SME \cref{eq:sme,eq:Mdiff}. A previous measurement aligns the angle of the heterodyne detection with the phase of $\alpha_2$, so we take $\alpha_2\in\mathbb{R}$ and consider only the $X$ quadrature signal (we just ignore the other measured signal). This is modelled by homodyne detection with halved rate $\kappa_1/2$, resulting in a single binned signal $\{I_0,\dots,I_N\}$ defined by \cref{eq:dYt,eq:Ik-diff}:
\begin{align}\label{eq:ex3-dYt}
    \begin{split}
        \dYt & = \sqrt{\eta\kappa_1/2}\,\Tr{(a+a^\dag)\rho_t}\dt + \dWt,    \\
        I_k  & = \frac{G}{\Delta t}\int_{k\Delta t}^{(k+1)\Delta t} \dYt.
    \end{split}
\end{align}
We assume that the system has reached its steady state $\rho_\infty$, which is also implicitly fitted since it depends on the parameters. We choose parameter values close to those of the experiment \cite{berdou2023one}, i.e. $\kappa_2\ll \kappa_1$ and a low measurement efficiency $\eta=0.1$.

The amplitude of $\alpha_2$ can be controlled by an external drive. There are three main regimes for the system steady state as $\alpha_2$ is increased:
\begin{itemize}
    \item For ${\alpha_2^2\ll\kappa_1/(2\kappa_2)}$, the steady state remains close to vacuum $\rho_\infty\approx\ket0\bra0$. The measured signal $I_k$ is just filtered white noise.
    \item For ${\alpha_2^2\approx\kappa_1/(2\kappa_2)}$ the system enters an interesting regime where the vacuum starts splitting into two coherent states, which are not yet fully separated in phase space.
    \item For ${\alpha_2^2\gg\kappa_1/(2\kappa_2)}$, the cavity state is pinned to one of the two coherent states $\ket{\pm\,\alpha}$ with average photon number ${\alpha^2=\alpha_2^2-\kappa_1/(2\kappa_2)}$. Sometimes, the state suddenly jumps from one coherent state to the other. These jumps can be observed individually on a single stochastic trajectory, as the measured signal $I_k$ flips from being centered at $\pm\, G\sqrt{2\eta\kappa_1}\alpha$ to its opposite value.
\end{itemize}
In this last regime, the characteristic flip time between the two coherent states $\ket{\pm\,\alpha}$ increases exponentially with $\alpha^2$. The goal of the experiment was to demonstrate that the engineered dissipation \cref{eq:HL1L2} allows to reach macroscopic flip times, up to a few hundred seconds, with a reasonably low number of photons $\alpha^2 < 100$.

\begin{figure*}[ht]
    \includegraphics{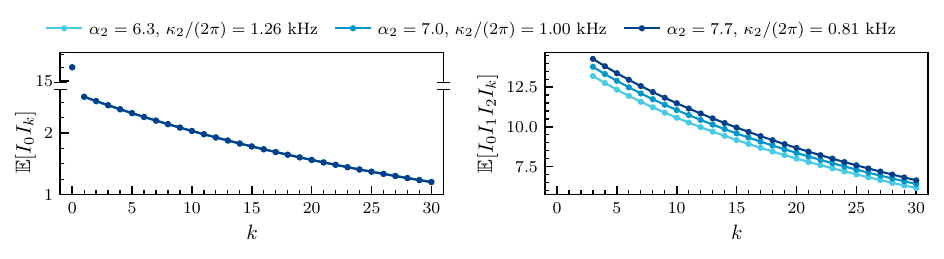}
    \vspace{-0.4cm}
    \caption{\textbf{An example of low-order degeneracy.} This figure shows for three different combinations of $\alpha_2$ and $\kappa_2$ (blue lines) the two-point correlation functions $\E{I_0I_k}$ for $k\geq0$ (left panel) and the four-point correlation functions $\E{I_0I_1I_2I_k}$ for $k\geq3$ (right panel). This highlights the system low-order degeneracy: the parameters $\alpha_2$ and $\kappa_2$ can't be estimated from the two-point correlation functions alone, because the different curves are superimposed. Fortunately, this is not a fundamental non-identifiability, as the degeneracy is lifted by the four-point correlation functions.}
    \label{fig:ex3/degeneracy}
\end{figure*}

\textbf{Parameters to be estimated} -- There are five parameters to be estimated experimentally: $\kappa_1$, $\kappa_2$, $\alpha_2$, $G$ and $\eta$. The gain was estimated by setting $\alpha_2=0$ and fitting the autocorrelation of the measured signal when the system remains in vacuum (see \cref{app:gain} for more details). We set $G=1.0$ for convenience. The parameter $\kappa_1$ was characterised independently. Our goal is to estimate the values of the remaining three parameters $\btheta=(\kappa_2, \alpha_2, \eta)$.

As explained in \cite{berdou2023one}, these parameters are particularly difficult to estimate independently using standard measurement techniques. For this experiment, reliably characterising the photon number $\alpha^2$, which depends on $\kappa_2$ and $\alpha_2$, was crucial to support the claim that macroscopic flip times were observed for states with a few tens of photons. A specifically designed sequence of experiments, including the estimation of the variance of the measured signal $\mathbb{E}[I_k^2]$ for various values of $\alpha_2$, ultimately allowed to characterise these three parameters. However, the resulting uncertainty for each estimate was greater than $10\,\%$. To gain confidence in this characterisation, the estimated order of magnitude for the quantum efficiency $\eta$ was independently verified by fabricating, cooling down, and measuring another superconducting chip, which was specifically conceived for this purpose.

Fitting the statistics of the measured signal already proved crucial in characterising the experiment, but only the signal variance was included in the fit. Importantly, our method leverages temporal information from correlation functions at non-coincident times. We discuss why the system was so challenging to characterise, and show how our approach could have been used as a simple alternative to estimate these three parameters with good precision.

\textbf{Low-order degeneracy} -- Interestingly, this system has a low-order degeneracy: for a fixed value of $\alpha_2$, it is not possible to independently estimate $\kappa_2$ and $\alpha_2$ from the one-, two- and three-point correlation functions.

First, any odd-order correlation function of the measured signal is null. This is a consequence of an interesting symmetry of the dynamics. If we introduce the parity superoperator $\Pcal$ defined by ${\Pcal(\rho)=e^{i\pi a^\dag a}\rho e^{-i\pi a^\dag a}}$, we observe that (i) it commutes with the Liouvillian $[\Lcal, \Pcal]=0$, and (ii) it anticommutes with the signal correlation superoperator $\{\Ccal_{L_1},\Pcal\}=0$. As a result, the stochastic process $\dYt$ is \emph{symmetric}: $\dYt$ and $-\,\dYt$ have the same law, or equivalently, all odd-order correlation functions are null. This observation can be generalised to any diffusive continuous measurement that satisfies conditions (i) and (ii) for some symmetry superoperator such as $\Pcal$. We give a straightforward proof of this result in \cref{app:symmetry}, using the correlation function formulae presented in \cref{sec:theory}.

The only information available in the signal statistics is thus to be found in the even-order correlation functions. However, the two-point correlation function is not sufficient to estimate the system parameters, because there is a low-order degeneracy between $\kappa_2$ and $\alpha_2$. We observe numerically that different combinations of values lead to almost indistinguishable two-point functions. The values of $\kappa_2$ and $\alpha_2$ that give the same two-point functions correspond to the same steady state photon number ${\alpha^2=\Tr{a^\dag a\,\rho_\infty}=f(\kappa_2, \alpha_2)}$, where $f$ is computed numerically. So albeit we cannot fully characterise the system with the two-point functions, they already give access to one of the sought-after quantities: the average photon number.

Let us see how to push the characterisation further to fully characterise $\alpha_2$ and $\kappa_2$. To start, we set ourselves in the intermediate regime where ${\alpha_2^2\approx\kappa_1/(2\kappa_2)}$. Note that even without knowing the precise value of $\alpha_2$, the transition between the three regimes can be witnessed experimentally, which allows to set a value of $\alpha_2$ \emph{somewhere} in this second regime. We propose two ways to lift the degeneracy. The first is to leverage the four-point correlation functions, which independently identifies $\kappa_2$ and $\alpha_2$. \Cref{fig:ex3/degeneracy} illustrates the low-order two-point degeneracy and how it is lifted by the four-point functions. However, estimating high-order correlation functions with precision is usually costly in terms of measurement time (see \cref{app:uncertainties}), especially when the efficiency is low. A second solution is to jointly fit the two-point functions for different configurations of the system parameters. As explained before, $\alpha_2$ is proportional to a controllable drive, so by scaling the drive amplitude the system can be measured with two different values $\alpha_2^a$ and $\alpha_2^b$, linked by a known factor. We now demonstrate how this allows to complete the system characterisation with good precision.

\begin{figure}[htb]
    \centering
    \includegraphics{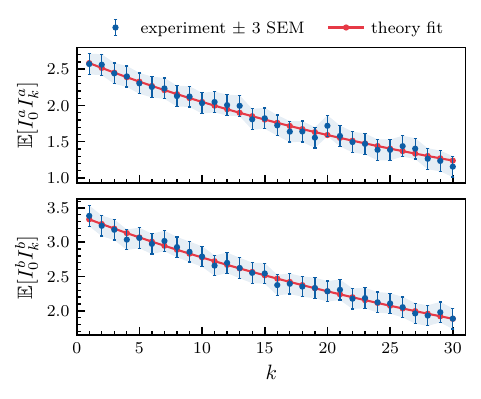}
    \vspace{-0.4cm}
    \caption{\textbf{Joint fit of the homodyne signal two-point correlation functions in two different configurations.} Experimental estimates (synthetic data, blue points) of the binned signal two-point correlation functions for two configurations (top panel: configuration $a$ with $\alpha_2^a=7.0$, bottom panel: configuration $b$ with $\alpha_2^b=7.14$) are simultaneously fitted by a theoretical model (red line). The blue error bars indicate a range of plus or minus three times the SEM of the experimental estimates.}
    \label{fig:ex3/fit}
\end{figure}

\textbf{Results} -- To model experimental data, we simulate ${\nexp=10^5}$ stochastic trajectories for two configurations $a$ and $b$ of the system, with respective values for the amplitude $\alpha_2^a=7.0=\alpha_2$ and $\alpha_2^b=1.02\,\alpha_2^a=7.14$. For each trajectory (labelled $j$), we average the measured signals over $31$ time bins of duration $\Delta t=1/(2\kappa_1)$, to obtain two binned signals, one for each configuration: $\{I_0^{a, (j)},\dots,I_{30}^{a, (j)}\}$ and $\{I_0^{b, (j)},\dots,I_{30}^{b,(j)}\}$. We estimate the two-point correlation functions of each signal for non-coincident time bins $\E{I_0I_k}\ (1\leq k\leq 30)$ by averaging over the $10^5$ trajectories. We then jointly fit the estimated correlation functions $\hat{\mathbb{E}}[I_0^{a/b}I_k^{a/b}]$ with a least-squares method. For this example, we fit the binned signal correlation functions using the sharp signal formula approximation (see end of \cref{sec:numerics}). We verified with the binned signal formula that the error resulting from this approximation is significantly smaller than the error caused by the statistical fluctuations of the experimental estimates. The three parameters $\btheta=(\kappa_2, \alpha_2, \eta)$ can be estimated simultaneously with good precision from these data. The result of the fit is shown in \cref{fig:ex3/fit} and the estimated parameter values are summarised in \cref{table:ex3}. \Cref{app:ex3} gives further details on the system, the simulated data and the fitting procedure.

\begin{table}[htb]
    \centering
    \begin{tabular}{|c|l|l|l|}
        \hline
                    & \textbf{Parameter}   & \textbf{Value}    & \textbf{Estimated} \\\hline\hline
        $\kappa_1$ & $1$-ph. loss rate & $100\ \mathrm{kHz}$ & \hspace{2.65em} --  \\
        $\kappa_2$ & $2$-ph. loss rate & $1\ \mathrm{kHz}$ & $1.007\ \pm\ 0.004\ \mathrm{kHz}$        \\
        $\alpha_2$ & amplitude            & $7$             & $7.002\ \pm\ 0.005$                     \\
        $\eta$     & efficiency           & $0.1$             & $0.095\ \pm\ 0.002$                     \\\hline
    \end{tabular}
    \caption{\textbf{Estimated parameters.} Result of the least-squares fit to simultaneously estimate the three parameters $\btheta=(\kappa_2, \alpha_2, \eta)$. The values are given with a factor of $1/(2\pi)$ where applicable. The standard deviation for each parameter is estimated by subsampling over $\nsubset=10$ subsets.}
    \label{table:ex3}
\end{table}

Jointly fitting data across different system configurations, as illustrated in this example by varying the drive controlling $\alpha_2$, is generally an effective way to lift any degeneracies (see \cref{app:degeneracy} on parameter degeneracy for more details). The proposed method provides a practical and straightforward approach to lifting the low-order degeneracy and estimating the remaining parameters with high precision, thereby completing the system characterisation.

\section{Discussion}\label{sec:discussion}

We now compare the method with the two other main approaches for estimating parameters from continuous quantum measurements: Bayesian inference and machine learning. We also discuss the main advantages of our approach in more detail, as well as its limitations.

\subsection{Comparison with existing methods}

\textbf{Bayesian inference} -- The main idea of this approach is to find the parameters that maximise the likelihood of the observed data. The likelihood can be computed as the trace of the solution to the linear SME, which is solved using the measured signal. To find the maximum likelihood estimate, early work discretised the parameter space to compute the likelihood at each point on a grid \cite{mabuchi1996dynamical,gambetta2001state}. The problem was then formulated as quantum filtering, by extending the Hilbert space to embed the discretised parameter space \cite{chase2009single,negretti2013estimation,six2015parameter,bompais2022parameter}. Some papers proposed to reduce the numerical cost by sampling the likelihood with stochastic samplers \cite{gammelmark2013bayesian, ralph2017multiparameter}, and recently a novel algorithm was proposed in the online setting \cite{clausen2024online} by approximating the gradient of the likelihood, to iteratively update the parameter.

This approach is optimal and has well established theoretical guarantees. In some cases it gives access to the full posterior distribution of the parameters, rather than just the maximum likelihood value. However, it is particularly numerically expensive, because (i) it requires evolving a separate SME for each recorded trajectory, and (ii) the cost of sampling the likelihood scales exponentially with the number of parameters, when discretised on a grid. In practice, the approach is hardly applicable to estimate more than a few parameters in a system larger than a qubit. Furthermore, it does not take into account arbitrary filtering: these works assume that the digitisation time $\Delta t$ is negligible compared to the timescales of the dynamics. Finally, the method gives very little insight into how the estimated parameters depend on the observed data, making it difficult to interpret and debug.

\textbf{Machine learning} -- Approaches based on machine learning are more recent. Some consider a \emph{supervised} setting, by training an artificial neural network to learn the inverse problem of mapping an ensemble of measured signals to a set of parameters \cite{greplova2017quantum,khanahmadi2021time,rinaldi2024parameter}. These neural networks are trained on synthetic data by simulating many SME trajectories for known values of the parameters. Other approaches consider an \emph{unsupervised} setting, by training an artificial neural network in-situ, using the physical device directly, to learn the system dynamics by best predicting the measurement outcomes \cite{krastanov2020unboxing, genois2021quantum}. The Hamiltonian or parameters are found by enforcing the physical structure of the SME into the neural network. A recent work addresses both settings with the same architecture \cite{tucker2024hamiltonian}.

These approaches are usually quite general: any signal post-processing can be included and they can also be applied to non-Markovian dynamics. In the supervised setting, adding artificial noise to the training data makes the model robust to potentially imperfectly modelled uncertainties or experimental noise. Once trained, the inference of the supervised models for a two-level system was shown to be especially time and memory efficient. This is particularly relevant when performing real-time estimation on a parameter that varies on a very small timescale, e.g. on the order of a few milliseconds. However, for each different system or set of parameters to be estimated, a specific model must be trained by sampling the SME to build a large training set, adapting the model architecture to the problem constraints, and testing and validating on synthetic data. This approach has yet to be demonstrated for large systems and multiple digits precision on the estimated parameters. In contrast, in the unsupervised setting, the model is trained directly online on experimental data, but then a large number of experimental runs are required to estimate the parameters. Finally, both supervised and unsupervised approaches share the interpretability drawback with the Bayesian inference approach: it is not easy to understand how the parameters are fitted from the data.

\subsection{Advantages}

The proposed approach is based on an explicit formula to fit the measured data, which offers several advantages.

\textbf{Generality} -- Let us first emphasise the generality of the proposed protocol. It is applicable to any continuously measured system modelled by a SME, of jump and/or diffusive type. Common detector imperfections, such as efficiency and dark count rate are natively taken into account, as well as arbitrary signal filtering and digitisation. Importantly, there is no restriction on the digitisation time $\Delta t$, which can be arbitrarily long. This is crucial, for example, in superconducting circuits, where the fast dynamics challenge the common assumption that the digitisation time is negligible.

\textbf{Simplicity} -- Another advantage of the approach is its simplicity, both conceptually, experimentally and numerically. Conceptually, the theoretical formula models the measured experimental data directly, without the need to reconstruct intermediate quantities such as the quantum state. Experimentally, the processing is straightforward, involving only the calculation of averages of the measured signal. Numerically, the fitting procedure is easily implemented in a dozen lines of code by combining ODE solvers and least-squares fitting routines that are readily available in general scientific computing libraries.

\textbf{Efficiency} -- The method never relies on numerically solving the SME for many trajectories, it only involves solving a few modified Lindblad master equations. As a result, it is suitable for large Hilbert space dimensions: the procedure easily runs on a CPU for typical bosonic problems. Moreover, the numerical cost scales only linearly with the number of parameters (it is even constant when backward differentiation is used to compute the gradient), making it possible to estimate dozens of parameters simultaneously.

\textbf{Interpretability} -- The interpretability of the method is its distinguishing feature. This is crucial when working with a new system, or when trying to understand new physics. Firstly, an experimentalist can develop an intuition about how each parameter uniquely affects the fitted correlation functions. Secondly, the goodness of the fit can be assessed visually, which helps to build confidence in the fitted parameters and validate the correctness of the model. To take this further, statistical tests such as the Chi-square tests can be used to determine whether the fitted model adequately explains the observed data, or whether there are systematic deviations. This method can therefore also be used as a tool to validate or reject a candidate theoretical model against experimental data.

These four main advantages make the approach suitable for everyday use as a new tool to characterise quantum systems. They also make it a relevant choice for automating the characterisation phase.

\subsection{Limitations}

We identify four main limitations.

\textbf{Prior model knowledge} -- The main limitation of our approach is the need for a parameterised model of the system. Note, however, that if the model is missing an important term, this will typically result in an incorrect fit rather than a biased fit, as each parameter generally has a unique influence on the signal correlation functions. This can be diagnosed if the data are poorly fitted by the chosen model, unless there is an unfortunate degeneracy between the missing term and one of the existing terms. If the model is uncertain, solutions include (i) performing model identification by adding various additional terms (such as cross-Kerr, drive line delays, parameterised filter functions, etc.), (ii) embedding robustness by finding the best fitting parameters for a range of added noise or uncertainties, or (iii) for small systems such as a qubit, fitting the full Hamiltonian.

\textbf{Optimality} -- The proposed scheme does not make optimal use of all the available information. However, although the optimal solution based on Bayesian inference has been known for some time, it has a number of drawbacks that make it impractical and not often used in practice, as discussed in the previous subsection.

\textbf{Theoretical guarantees} -- There are no firm theoretical guarantees about, for example, the bias of the estimator and the convergence of the method. This paper is a somewhat \emph{empirical} demonstration of the utility of the method, but it still lacks a rigorous theoretical framework. Further theoretical work includes establishing conditions for parameter identifiability, comparing with the optimal Bayesian scheme, finding an explicit formula for the parameters variance as a function of the number of trajectories, and proving the convergence and unbiasedness of the estimator.

\textbf{Posterior distribution} -- Another limitation of the method is that it only provides a single estimate for the parameters, rather than the full posterior distribution. In practice, and as explained, the variance can be computed by simply subsampling the data or bootstrapping the procedure, which is sufficient for many characterisation tasks.

\section{Conclusion}\label{sec:conclusion}

Continuously measured signals provide a wealth of information about the dynamics of a quantum system. In this paper, we have presented and demonstrated a simple method to leverage this information to determine the system parameters, by fitting correlation functions of the measured signals.

The approach stands out for its generality, its simplicity and its efficiency, without sacrificing interpretability. The method is broadly applicable to any system modelled by a jump or diffusive SME, it allows the simultaneous estimation of multiple parameters in Hilbert spaces of large dimensions, and it takes into account typical detector imperfections as well as arbitrary signal filtering and digitisation. It is easy to apply experimentally, and the numerics are simple to implement. Finally, much intuition can be developed about the fitting procedure, allowing for easier debugging and model validation.

While our examples have focused on systems with a limited number of experimental control knobs, the effectiveness of the method can be further enhanced by measuring data in different regimes of the system dynamics, by adjusting these knobs at will. Jointly fitting signal statistics across different system configurations, for example starting from different initial states, or varying the amplitude and phase of linear drives, is a particularly efficient way to lift any degeneracies, improve the precision of the estimated parameters and validate the model correctness. The method also lends itself to tracking parameter drift in an online setting, by repeatedly recomputing correlation functions on newly acquired data. Finally, the procedure is well-suited to automate the calibration process of large quantum systems, as it allows the simultaneous estimation of many parameters with a simple measurement and data post-processing.

In conclusion, this characterisation method is a new powerful tool for researchers and engineers working with increasingly complex quantum systems.

\section*{Acknowledgements}

This project has received funding from the Plan France 2030 through the project ANR-22-PETQ-0006, as well as from the European Research Council (ERC) under the European Union’s Horizon 2020 research and innovation programme (grant agreement No. 884762), and also from the ERC QFT.zip project (grant agreement No. 101040260).

\printbibliography

@article{krinner2022realizing,
  title   = {Realizing repeated quantum error correction in a distance-three surface code},
  author  = {Krinner,  Sebastian and Lacroix,  Nathan and Remm,  Ants and Di Paolo,  Agustin and Genois,  Elie and Leroux,  Catherine and Hellings,  Christoph and Lazar,  Stefania and Swiadek,  Francois and Herrmann,  Johannes and Norris,  Graham J. and Andersen,  Christian Kraglund and M\"{u}ller,  Markus and Blais,  Alexandre and Eichler,  Christopher and Wallraff,  Andreas},
  year    = {2022},
  month   = may,
  journal = {Nature},
  volume  = {605},
  number  = {7911},
  pages   = {669--674},
  url     = {http://dx.doi.org/10.1038/s41586-022-04566-8},
  doi     = {10.1038/s41586-022-04566-8},
  note    = {clean}
}

@article{mirrahimi2014dynamically,
  title   = {Dynamically protected cat-qubits: a new paradigm for universal quantum computation},
  author  = {Mirrahimi,  Mazyar and Leghtas,  Zaki and Albert,  Victor V and Touzard,  Steven and Schoelkopf,  Robert J and Jiang,  Liang and Devoret,  Michel H},
  year    = {2014},
  month   = apr,
  journal = {New Journal of Physics},
  volume  = {16},
  number  = {4},
  pages   = {045014},
  url     = {http://dx.doi.org/10.1088/1367-2630/16/4/045014},
  doi     = {10.1088/1367-2630/16/4/045014},
  note    = {clean}
}

@article{berdou2023one,
  title   = {One Hundred Second Bit-Flip Time in a Two-Photon Dissipative OsciPllator},
  author  = {Berdou,  C. and Murani,  A. and Réglade,  U. and Smith,  W.C. and Villiers,  M. and Palomo,  J. and Rosticher,  M. and Denis,  A. and Morfin,  P. and Delbecq,  M. and Kontos,  T. and Pankratova,  N. and Rautschke,  F. and Peronnin,  T. and Sellem,  L.-A. and Rouchon,  P. and Sarlette,  A. and Mirrahimi,  M. and Campagne-Ibarcq,  P. and Jezouin,  S. and Lescanne,  R. and Leghtas,  Z.},
  year    = {2023},
  month   = jun,
  journal = {PRX Quantum},
  volume  = {4},
  number  = {2},
  url     = {http://dx.doi.org/10.1103/PRXQuantum.4.020350},
  doi     = {10.1103/prxquantum.4.020350},
  note    = {clean}
}

@article{tilloy2018exact,
  title   = {Exact signal correlators in continuous quantum measurements},
  author  = {Tilloy,  Antoine},
  year    = {2018},
  month   = jul,
  journal = {Physical Review A},
  volume  = {98},
  number  = {1},
  url     = {https://doi.org/10.1103/physreva.98.010104},
  doi     = {10.1103/physreva.98.010104},
  note    = {clean}
}

@article{rouchon2015efficient,
  title   = {Efficient quantum filtering for quantum feedback control},
  author  = {Rouchon,  Pierre and Ralph,  Jason F.},
  year    = {2015},
  month   = jan,
  journal = {Physical Review A},
  volume  = {91},
  number  = {1},
  pages   = {012118},
  url     = {http://dx.doi.org/10.1103/PhysRevA.91.012118},
  doi     = {10.1103/physreva.91.012118},
  note    = {clean}
}

@article{jacobs2006straightforward,
  title   = {A straightforward introduction to continuous quantum measurement},
  author  = {Jacobs,  Kurt and Steck,  Daniel A.},
  year    = {2006},
  month   = sep,
  journal = {Contemporary Physics},
  volume  = {47},
  number  = {5},
  pages   = {279--303},
  url     = {https://doi.org/10.1080/00107510601101934},
  doi     = {10.1080/00107510601101934},
  note    = {clean}
}

@inproceedings{six2015parameter,
  title     = {Parameter estimation from measurements along quantum trajectories},
  author    = {Six,  P. and Campagne-Ibarcq,  Ph. and Bretheau,  L. and Huard,  B. and Rouchon,  P.},
  year      = {2015},
  month     = dec,
  booktitle = {2015 54th IEEE Conference on Decision and Control (CDC)},
  publisher = {IEEE},
  url       = {https://doi.org/10.1109/cdc.2015.7403443},
  doi       = {10.1109/cdc.2015.7403443},
  note      = {clean}
}

@book{nielsen2010quantum,
  title     = {Quantum Computation and Quantum Information: 10th Anniversary Edition},
  author    = {Nielsen,  Michael A. and Chuang,  Isaac L.},
  year      = {2010},
  publisher = {Cambridge University Press},
  url       = {http://dx.doi.org/10.1017/CBO9780511976667},
  doi       = {10.1017/cbo9780511976667},
  isbn      = {9780511976667},
  note      = {clean}
}

@article{acharya2023suppressing,
  title   = {Suppressing quantum errors by scaling a surface code logical qubit},
  author  = {Acharya,  Rajeev and Aleiner,  Igor and Allen,  Richard and Andersen,  Trond I. and Ansmann,  Markus and Arute,  Frank and Arya,  Kunal and Asfaw,  Abraham and Atalaya,  Juan and Babbush,  Ryan and Bacon,  Dave and Bardin,  Joseph C. and Basso,  Joao and Bengtsson,  Andreas and Boixo,  Sergio and Bortoli,  Gina and Bourassa,  Alexandre and Bovaird,  Jenna and Brill,  Leon and Broughton,  Michael and Buckley,  Bob B. and Buell,  David A. and Burger,  Tim and Burkett,  Brian and Bushnell,  Nicholas and Chen,  Yu and Chen,  Zijun and Chiaro,  Ben and Cogan,  Josh and Collins,  Roberto and Conner,  Paul and Courtney,  William and Crook,  Alexander L. and Curtin,  Ben and Debroy,  Dripto M. and Del Toro Barba,  Alexander and Demura,  Sean and Dunsworth,  Andrew and Eppens,  Daniel and Erickson,  Catherine and Faoro,  Lara and Farhi,  Edward and Fatemi,  Reza and Flores Burgos,  Leslie and Forati,  Ebrahim and Fowler,  Austin G. and Foxen,  Brooks and Giang,  William and Gidney,  Craig and Gilboa,  Dar and Giustina,  Marissa and Grajales Dau,  Alejandro and Gross,  Jonathan A. and Habegger,  Steve and Hamilton,  Michael C. and Harrigan,  Matthew P. and Harrington,  Sean D. and Higgott,  Oscar and Hilton,  Jeremy and Hoffmann,  Markus and Hong,  Sabrina and Huang,  Trent and Huff,  Ashley and Huggins,  William J. and Ioffe,  Lev B. and Isakov,  Sergei V. and Iveland,  Justin and Jeffrey,  Evan and Jiang,  Zhang and Jones,  Cody and Juhas,  Pavol and Kafri,  Dvir and Kechedzhi,  Kostyantyn and Kelly,  Julian and Khattar,  Tanuj and Khezri,  Mostafa and Kieferová,  Mária and Kim,  Seon and Kitaev,  Alexei and Klimov,  Paul V. and Klots,  Andrey R. and Korotkov,  Alexander N. and Kostritsa,  Fedor and Kreikebaum,  John Mark and Landhuis,  David and Laptev,  Pavel and Lau,  Kim-Ming and Laws,  Lily and Lee,  Joonho and Lee,  Kenny and Lester,  Brian J. and Lill,  Alexander and Liu,  Wayne and Locharla,  Aditya and Lucero,  Erik and Malone,  Fionn D. and Marshall,  Jeffrey and Martin,  Orion and McClean,  Jarrod R. and McCourt,  Trevor and McEwen,  Matt and Megrant,  Anthony and Meurer Costa,  Bernardo and Mi,  Xiao and Miao,  Kevin C. and Mohseni,  Masoud and Montazeri,  Shirin and Morvan,  Alexis and Mount,  Emily and Mruczkiewicz,  Wojciech and Naaman,  Ofer and Neeley,  Matthew and Neill,  Charles and Nersisyan,  Ani and Neven,  Hartmut and Newman,  Michael and Ng,  Jiun How and Nguyen,  Anthony and Nguyen,  Murray and Niu,  Murphy Yuezhen and O’Brien,  Thomas E. and Opremcak,  Alex and Platt,  John and Petukhov,  Andre and Potter,  Rebecca and Pryadko,  Leonid P. and Quintana,  Chris and Roushan,  Pedram and Rubin,  Nicholas C. and Saei,  Negar and Sank,  Daniel and Sankaragomathi,  Kannan and Satzinger,  Kevin J. and Schurkus,  Henry F. and Schuster,  Christopher and Shearn,  Michael J. and Shorter,  Aaron and Shvarts,  Vladimir and Skruzny,  Jindra and Smelyanskiy,  Vadim and Smith,  W. Clarke and Sterling,  George and Strain,  Doug and Szalay,  Marco and Torres,  Alfredo and Vidal,  Guifre and Villalonga,  Benjamin and Vollgraff Heidweiller,  Catherine and White,  Theodore and Xing,  Cheng and Yao,  Z. Jamie and Yeh,  Ping and Yoo,  Juhwan and Young,  Grayson and Zalcman,  Adam and Zhang,  Yaxing and Zhu,  Ningfeng},
  year    = {2023},
  month   = feb,
  journal = {Nature},
  volume  = {614},
  number  = {7949},
  pages   = {676–681},
  url     = {http://dx.doi.org/10.1038/s41586-022-05434-1},
  doi     = {10.1038/s41586-022-05434-1},
  note    = {clean}
}

@article{lieu2020symmetry,
  title   = {Symmetry Breaking and Error Correction in Open Quantum Systems},
  author  = {Lieu,  Simon and Belyansky,  Ron and Young,  Jeremy T. and Lundgren,  Rex and Albert,  Victor V. and Gorshkov,  Alexey V.},
  year    = {2020},
  month   = dec,
  journal = {Physical Review Letters},
  volume  = {125},
  number  = {24},
  url     = {https://doi.org/10.1103/physrevlett.125.240405},
  doi     = {10.1103/physrevlett.125.240405},
  note    = {clean}
}

@book{wiseman2009quantum,
  title     = {Quantum Measurement and Control},
  author    = {Wiseman,  Howard M. and Milburn,  Gerard J.},
  year      = {2009},
  month     = nov,
  publisher = {Cambridge University Press},
  url       = {https://doi.org/10.1017/cbo9780511813948},
  doi       = {10.1017/cbo9780511813948},
  isbn      = {9781107424159},
  note      = {clean}
}

@article{rouchon2022tutorial,
  title   = {A tutorial introduction to quantum stochastic master equations based on the qubit/photon system},
  author  = {Rouchon,  Pierre},
  year    = {2022},
  journal = {Annual Reviews in Control},
  volume  = {54},
  pages   = {252--261},
  url     = {https://doi.org/10.1016/j.arcontrol.2022.09.006},
  doi     = {10.1016/j.arcontrol.2022.09.006},
  note    = {clean}
}

@book{barchielli2009quantum,
  title     = {Quantum Trajectories and Measurements in Continuous Time: The Diffusive Case},
  author    = {Barchielli,  Alberto and Gregoratti,  Matteo},
  year      = {2009},
  publisher = {Springer},
  url       = {https://doi.org/10.1007/978-3-642-01298-3},
  doi       = {10.1007/978-3-642-01298-3},
  note      = {clean}
}

@article{atalaya2018multitime,
  title   = {Multitime correlators in continuous measurement of qubit observables},
  author  = {Atalaya,  Juan and Hacohen-Gourgy,  Shay and Martin,  Leigh S. and Siddiqi,  Irfan and Korotkov,  Alexander N.},
  year    = {2018},
  month   = feb,
  journal = {Physical Review A},
  volume  = {97},
  number  = {2},
  url     = {https://doi.org/10.1103/physreva.97.020104},
  doi     = {10.1103/physreva.97.020104},
  note    = {clean}
}

@article{diosi2016structural,
  title   = {Structural features of sequential weak measurements},
  author  = {Diósi,  Lajos},
  year    = {2016},
  month   = jul,
  journal = {Physical Review A},
  volume  = {94},
  number  = {1},
  url     = {https://doi.org/10.1103/physreva.94.010103},
  doi     = {10.1103/physreva.94.010103},
  note    = {clean}
}

@article{foroozani2016correlations,
  title   = {Correlations of the Time Dependent Signal and the State of a Continuously Monitored Quantum System},
  author  = {Foroozani,  N. and Naghiloo,  M. and Tan,  D. and Mølmer,  K. and Murch,  K. W.},
  year    = {2016},
  month   = mar,
  journal = {Physical Review Letters},
  volume  = {116},
  number  = {11},
  url     = {https://doi.org/10.1103/physrevlett.116.110401},
  doi     = {10.1103/physrevlett.116.110401},
  note    = {clean}
}

@article{korotkov2001output,
  title   = {Output spectrum of a detector measuring quantum oscillations},
  author  = {Korotkov,  Alexander N.},
  year    = {2001},
  month   = feb,
  journal = {Physical Review B},
  volume  = {63},
  number  = {8},
  url     = {https://doi.org/10.1103/physrevb.63.085312},
  doi     = {10.1103/physrevb.63.085312},
  note    = {clean}
}

@article{genois2021quantum,
  title   = {Quantum-Tailored Machine-Learning Characterization of a Superconducting Qubit},
  author  = {Genois,  Élie and Gross,  Jonathan A. and Di Paolo,  Agustin and Stevenson,  Noah J. and Koolstra,  Gerwin and Hashim,  Akel and Siddiqi,  Irfan and Blais,  Alexandre},
  year    = {2021},
  month   = dec,
  journal = {PRX Quantum},
  volume  = {2},
  number  = {4},
  pages   = {040355},
  url     = {http://dx.doi.org/10.1103/PRXQuantum.2.040355},
  doi     = {10.1103/prxquantum.2.040355},
  note    = {clean}
}

@article{jordan2016anatomy,
  title   = {Anatomy of fluorescence: quantum trajectory statistics from continuously measuring spontaneous emission},
  author  = {Jordan,  Andrew N. and Chantasri,  Areeya and Rouchon,  Pierre and Huard,  Benjamin},
  year    = {2016},
  month   = may,
  journal = {Quantum Studies: Mathematics and Foundations},
  volume  = {3},
  number  = {3},
  pages   = {237--263},
  url     = {https://doi.org/10.1007/s40509-016-0075-9},
  doi     = {10.1007/s40509-016-0075-9},
  note    = {clean}
}

@article{landi2024current,
  title   = {Current Fluctuations in Open Quantum Systems: Bridging the Gap Between Quantum Continuous Measurements and Full Counting Statistics},
  author  = {Landi,  Gabriel T. and Kewming,  Michael J. and Mitchison,  Mark T. and Potts,  Patrick P.},
  year    = {2024},
  month   = apr,
  journal = {PRX Quantum},
  volume  = {5},
  number  = {2},
  url     = {http://dx.doi.org/10.1103/PRXQuantum.5.020201},
  doi     = {10.1103/prxquantum.5.020201},
  note    = {clean}
}

@article{flindt2010counting,
  title   = {Counting statistics of transport through Coulomb blockade nanostructures: High-order cumulants and non-Markovian effects},
  author  = {Flindt,  Christian and Novotný,  Tomáš and Braggio,  Alessandro and Jauho,  Antti-Pekka},
  year    = {2010},
  month   = oct,
  journal = {Physical Review B},
  volume  = {82},
  number  = {15},
  url     = {https://doi.org/10.1103/physrevb.82.155407},
  doi     = {10.1103/physrevb.82.155407},
  note    = {clean}
}

@article{hagele2018higher,
  title   = {Higher-order moments, cumulants, and spectra of continuous quantum noise measurements},
  author  = {H\"{a}gele,  Daniel and Schefczik,  Fabian},
  year    = {2018},
  month   = nov,
  journal = {Physical Review B},
  volume  = {98},
  number  = {20},
  url     = {https://doi.org/10.1103/physrevb.98.205143},
  doi     = {10.1103/physrevb.98.205143},
  note    = {clean}
}

@article{sifft2021quantum,
  title   = {Quantum polyspectra for modeling and evaluating quantum transport measurements: A unifying approach to the strong and weak measurement regime},
  author  = {Sifft,  M. and Kurzmann,  A. and Kerski,  J. and Schott,  R. and Ludwig,  A. and Wieck,  A. D. and Lorke,  A. and Geller,  M. and H\"{a}gele,  D.},
  year    = {2021},
  month   = aug,
  journal = {Physical Review Research},
  volume  = {3},
  number  = {3},
  url     = {https://doi.org/10.1103/physrevresearch.3.033123},
  doi     = {10.1103/physrevresearch.3.033123},
  note    = {clean}
}

@article{xu2015correlation,
  title   = {Correlation functions and conditioned quantum dynamics in photodetection theory},
  author  = {Xu,  Qing and Greplova,  Eliska and Julsgaard,  Brian and Klaus Mølmer},
  year    = {2015},
  month   = nov,
  journal = {Physica Scripta},
  volume  = {90},
  number  = {12},
  pages   = {128004},
  url     = {https://doi.org/10.1088/0031-8949/90/12/128004},
  doi     = {10.1088/0031-8949/90/12/128004},
  note    = {clean}
}

@article{barchielli1991measurements,
  title   = {Measurements continuous in time and a posteriori states in quantum mechanics},
  author  = {Barchielli,  A and Belavkin,  V P},
  year    = {1991},
  month   = apr,
  journal = {Journal of Physics A: Mathematical and General},
  volume  = {24},
  number  = {7},
  pages   = {1495--1514},
  url     = {https://doi.org/10.1088/0305-4470/24/7/022},
  doi     = {10.1088/0305-4470/24/7/022},
  note    = {clean}
}

@article{guilmin2023correlation,
  title   = {Correlation functions for realistic continuous quantum measurement},
  author  = {Guilmin,  Pierre and Rouchon,  Pierre and Tilloy,  Antoine},
  year    = {2023},
  journal = {IFAC-PapersOnLine},
  volume  = {56},
  number  = {2},
  pages   = {5164–5170},
  url     = {http://dx.doi.org/10.1016/j.ifacol.2023.10.110},
  doi     = {10.1016/j.ifacol.2023.10.110},
  note    = {clean}
}

@article{minganti2016exact,
  title   = {Exact results for Schrödinger cats in driven-dissipative systems and their feedback control},
  author  = {Minganti,  Fabrizio and Bartolo,  Nicola and Lolli,  Jared and Casteels,  Wim and Ciuti,  Cristiano},
  year    = {2016},
  month   = may,
  journal = {Scientific Reports},
  volume  = {6},
  number  = {1},
  url     = {https://doi.org/10.1038/srep26987},
  doi     = {10.1038/srep26987},
  note    = {clean}
}

@article{gammelmark2013bayesian,
  title   = {Bayesian parameter inference from continuously monitored quantum systems},
  author  = {Gammelmark,  Søren and Mølmer,  Klaus},
  year    = {2013},
  month   = mar,
  journal = {Physical Review A},
  volume  = {87},
  number  = {3},
  pages   = {032115},
  url     = {http://dx.doi.org/10.1103/PhysRevA.87.032115},
  doi     = {10.1103/physreva.87.032115},
  note    = {clean}
}

@article{rinaldi2024parameter,
  title   = {Parameter estimation from quantum-jump data using neural networks},
  author  = {Rinaldi,  Enrico and González Lastre,  Manuel and García Herreros,  Sergio and Ahmed,  Shahnawaz and Khanahmadi,  Maryam and Nori,  Franco and Sánchez Muñoz,  Carlos},
  year    = {2024},
  month   = apr,
  journal = {Quantum Science and Technology},
  volume  = {9},
  number  = {3},
  pages   = {035018},
  url     = {http://dx.doi.org/10.1088/2058-9565/ad3c68},
  doi     = {10.1088/2058-9565/ad3c68},
  note    = {clean}
}

@unpublished{guilmin2024dynamiqs,
  title  = {Dynamiqs: an open-source Python library for GPU-accelerated and differentiable simulation of quantum systems},
  author = {Pierre Guilmin and Ronan Gautier and Adrien Bocquet and {\'{E}}lie Genois},
  year   = {2024},
  url    = {https://github.com/dynamiqs/dynamiqs},
  note   = {In preparation}
}

@article{kidger2022on,
  title   = {On Neural Differential Equations},
  author  = {Kidger,  Patrick},
  year    = {2022},
  journal = {arXiv preprint},
  number  = {2202.02435},
  url     = {https://arxiv.org/abs/2202.02435},
  doi     = {10.48550/ARXIV.2202.02435},
  note    = {clean}
}

@software{jax2018github,
  title   = {{JAX}: composable transformations of {P}ython+{N}um{P}y programs},
  author  = {James Bradbury and Roy Frostig and Peter Hawkins and Matthew James Johnson and Chris Leary and Dougal Maclaurin and George Necula and Adam Paszke and Jake Vander{P}las and Skye Wanderman-{M}ilne and Qiao Zhang},
  year    = {2018},
  version = {0.3.13},
  url     = {http://github.com/google/jax},
  note    = {clean}
}

@article{virtanen2020scipy,
  title   = {SciPy 1.0: fundamental algorithms for scientific computing in Python},
  author  = {Virtanen,  Pauli and Gommers,  Ralf and Oliphant,  Travis E. and Haberland,  Matt and Reddy,  Tyler and Cournapeau,  David and Burovski,  Evgeni and Peterson,  Pearu and Weckesser,  Warren and Bright,  Jonathan and van der Walt,  Stéfan J. and Brett,  Matthew and Wilson,  Joshua and Millman,  K. Jarrod and Mayorov,  Nikolay and Nelson,  Andrew R. J. and Jones,  Eric and Kern,  Robert and Larson,  Eric and Carey,  C J and Polat,  İlhan and Feng,  Yu and Moore,  Eric W. and VanderPlas,  Jake and Laxalde,  Denis and Perktold,  Josef and Cimrman,  Robert and Henriksen,  Ian and Quintero,  E. A. and Harris,  Charles R. and Archibald,  Anne M. and Ribeiro,  Ant\^onio H. and Pedregosa,  Fabian and van Mulbregt,  Paul and Vijaykumar,  Aditya and Bardelli,  Alessandro Pietro and Rothberg,  Alex and Hilboll,  Andreas and Kloeckner,  Andreas and Scopatz,  Anthony and Lee,  Antony and Rokem,  Ariel and Woods,  C. Nathan and Fulton,  Chad and Masson,  Charles and H\"{a}ggstr\"{o}m,  Christian and Fitzgerald,  Clark and Nicholson,  David A. and Hagen,  David R. and Pasechnik,  Dmitrii V. and Olivetti,  Emanuele and Martin,  Eric and Wieser,  Eric and Silva,  Fabrice and Lenders,  Felix and Wilhelm,  Florian and Young,  G. and Price,  Gavin A. and Ingold,  Gert-Ludwig and Allen,  Gregory E. and Lee,  Gregory R. and Audren,  Hervé and Probst,  Irvin and Dietrich,  J\"{o}rg P. and Silterra,  Jacob and Webber,  James T and Slavič,  Janko and Nothman,  Joel and Buchner,  Johannes and Kulick,  Johannes and Sch\"{o}nberger,  Johannes L. and de Miranda Cardoso,  José Vinícius and Reimer,  Joscha and Harrington,  Joseph and Rodríguez,  Juan Luis Cano and Nunez-Iglesias,  Juan and Kuczynski,  Justin and Tritz,  Kevin and Thoma,  Martin and Newville,  Matthew and K\"{u}mmerer,  Matthias and Bolingbroke,  Maximilian and Tartre,  Michael and Pak,  Mikhail and Smith,  Nathaniel J. and Nowaczyk,  Nikolai and Shebanov,  Nikolay and Pavlyk,  Oleksandr and Brodtkorb,  Per A. and Lee,  Perry and McGibbon,  Robert T. and Feldbauer,  Roman and Lewis,  Sam and Tygier,  Sam and Sievert,  Scott and Vigna,  Sebastiano and Peterson,  Stefan and More,  Surhud and Pudlik,  Tadeusz and Oshima,  Takuya and Pingel,  Thomas J. and Robitaille,  Thomas P. and Spura,  Thomas and Jones,  Thouis R. and Cera,  Tim and Leslie,  Tim and Zito,  Tiziano and Krauss,  Tom and Upadhyay,  Utkarsh and Halchenko,  Yaroslav O. and Vázquez-Baeza,  Yoshiki},
  year    = {2020},
  month   = feb,
  journal = {Nature Methods},
  volume  = {17},
  number  = {3},
  pages   = {261–272},
  url     = {http://dx.doi.org/10.1038/s41592-019-0686-2},
  doi     = {10.1038/s41592-019-0686-2},
  note    = {clean}
}

@article{rouchon2014models,
  title   = {Models and Feedback Stabilization of Open Quantum Systems},
  author  = {Rouchon,  Pierre},
  year    = {2014},
  journal = {arXiv preprint},
  number   = {1407.7810},
  url     = {https://arxiv.org/abs/1407.7810},
  doi     = {10.48550/ARXIV.1407.7810},
  note    = {clean}
}

@article{mabuchi1996dynamical,
  title   = {Dynamical identification of open quantum systems},
  author  = {Mabuchi,  H},
  year    = {1996},
  month   = dec,
  journal = {Quantum and Semiclassical Optics: Journal of the European Optical Society Part B},
  volume  = {8},
  number  = {6},
  pages   = {1103–1108},
  url     = {http://dx.doi.org/10.1088/1355-5111/8/6/002},
  doi     = {10.1088/1355-5111/8/6/002},
  note    = {clean}
}

@article{gambetta2001state,
  title   = {State and dynamical parameter estimation for open quantum systems},
  author  = {Gambetta,  Jay and Wiseman,  H. M.},
  year    = {2001},
  month   = sep,
  journal = {Physical Review A},
  volume  = {64},
  number  = {4},
  pages   = {042105},
  url     = {http://dx.doi.org/10.1103/PhysRevA.64.042105},
  doi     = {10.1103/physreva.64.042105},
  note    = {clean}
}

@article{chase2009single,
  title   = {Single-shot parameter estimation via continuous quantum measurement},
  author  = {Chase,  Bradley A. and Geremia,  J. M.},
  year    = {2009},
  month   = feb,
  journal = {Physical Review A},
  volume  = {79},
  number  = {2},
  pages   = {022314},
  url     = {http://dx.doi.org/10.1103/PhysRevA.79.022314},
  doi     = {10.1103/physreva.79.022314},
  note    = {clean}
}

@article{negretti2013estimation,
  title   = {Estimation of classical parameters via continuous probing of complementary quantum observables},
  author  = {Negretti,  A and Mølmer,  K},
  year    = {2013},
  month   = dec,
  journal = {New Journal of Physics},
  volume  = {15},
  number  = {12},
  pages   = {125002},
  url     = {http://dx.doi.org/10.1088/1367-2630/15/12/125002},
  doi     = {10.1088/1367-2630/15/12/125002},
  note    = {clean}
}

@article{greplova2017quantum,
  title   = {Quantum parameter estimation with a neural network},
  author  = {Greplova,  Eliska and Andersen,  Christian Kraglund and Mølmer,  Klaus},
  year    = {2017},
  journal = {arXiv preprint},
  number  = {1711.05238},
  url     = {https://arxiv.org/abs/1711.05238},
  doi     = {10.48550/ARXIV.1711.05238},
  note    = {clean}
}

@article{khanahmadi2021time,
  title   = {Time-dependent atomic magnetometry with a recurrent neural network},
  author  = {Khanahmadi,  Maryam and Mølmer,  Klaus},
  year    = {2021},
  month   = mar,
  journal = {Physical Review A},
  volume  = {103},
  number  = {3},
  pages   = {032406},
  url     = {http://dx.doi.org/10.1103/PhysRevA.103.032406},
  doi     = {10.1103/physreva.103.032406},
  note    = {clean}
}

@inproceedings{bompais2022parameter,
  title     = {Parameter Estimation for Quantum Trajectories: Convergence Result},
  author    = {Bompais,  Mael and Amini,  Nina H. and Pellegrini,  Clement},
  year      = {2022},
  month     = dec,
  booktitle = {2022 IEEE 61st Conference on Decision and Control (CDC)},
  publisher = {IEEE},
  pages     = {5161--5166},
  url       = {http://dx.doi.org/10.1109/CDC51059.2022.9992617},
  doi       = {10.1109/cdc51059.2022.9992617},
  note      = {clean}
}

@article{ralph2017multiparameter,
  title   = {Multiparameter estimation along quantum trajectories with sequential Monte Carlo methods},
  author  = {Ralph,  Jason F. and Maskell,  Simon and Jacobs,  Kurt},
  year    = {2017},
  month   = nov,
  journal = {Physical Review A},
  volume  = {96},
  number  = {5},
  pages   = {052306},
  url     = {http://dx.doi.org/10.1103/PhysRevA.96.052306},
  doi     = {10.1103/physreva.96.052306},
  note    = {clean}
}

@article{clausen2024online,
  title   = {Online Parameter Estimation for Continuously Monitored Quantum Systems},
  author  = {Clausen,  Henrik Glavind and Rouchon,  Pierre and Wisniewski,  Rafal},
  year    = {2024},
  journal = {IEEE Control Systems Letters},
  volume  = {8},
  pages   = {1247–-1252},
  url     = {http://dx.doi.org/10.1109/LCSYS.2024.3407608},
  doi     = {10.1109/lcsys.2024.3407608},
  note    = {clean}
}

@article{tucker2024hamiltonian,
  title   = {Hamiltonian Learning using Machine Learning Models Trained with Continuous Measurements},
  author  = {Tucker,  Kris and Rege,  Amit Kiran and Smith,  Conor and Monteleoni,  Claire and Albash,  Tameem},
  year    = {2024},
  journal = {arXiv preprint},
  number  = {2404.05526},
  url     = {https://arxiv.org/abs/2404.05526},
  doi     = {10.48550/ARXIV.2404.05526},
  note    = {clean}
}

@article{krastanov2020unboxing,
  title   = {Unboxing Quantum Black Box Models: Learning Non-Markovian Dynamics},
  author  = {Krastanov,  Stefan and Head-Marsden,  Kade and Zhou,  Sisi and Flammia,  Steven T. and Jiang,  Liang and Narang,  Prineha},
  year    = {2020},
  journal = {arXiv preprint},
  number  = {2009.03902},
  url     = {https://arxiv.org/abs/2009.03902},
  doi     = {10.48550/ARXIV.2009.03902},
  note    = {clean}
}

@article{sifft2024quantum,
  title   = {Quantum polyspectra approach to the dynamics of blinking quantum emitters at low photon rates without binning: Making every photon count},
  author  = {Sifft,  M. and Kurzmann,  A. and Kerski,  J. and Schott,  R. and Ludwig,  A. and Wieck,  A. D. and Lorke,  A. and Geller,  M. and H\"{a}gele,  D.},
  year    = {2024},
  month   = jun,
  journal = {Physical Review A},
  volume  = {109},
  number  = {6},
  pages   = {062210},
  url     = {http://dx.doi.org/10.1103/PhysRevA.109.062210},
  doi     = {10.1103/physreva.109.062210},
  note    = {clean}
}

@article{sifft2023random,
  title   = {Random-time quantum measurements},
  author  = {Sifft,  Markus and H\"{a}gele,  Daniel},
  year    = {2023},
  month   = may,
  journal = {Physical Review A},
  volume  = {107},
  number  = {5},
  pages   = {052203},
  url     = {http://dx.doi.org/10.1103/PhysRevA.107.052203},
  doi     = {10.1103/physreva.107.052203},
  note    = {clean}
}

@article{zhao2022realization,
  title   = {Realization of an error-correcting surface code with superconducting qubits},
  author  = {Zhao,  Youwei and Ye,  Yangsen and Huang,  He-Liang and Zhang,  Yiming and Wu,  Dachao and Guan,  Huijie and Zhu,  Qingling and Wei,  Zuolin and He,  Tan and Cao,  Sirui and Chen,  Fusheng and Chung,  Tung-Hsun and Deng,  Hui and Fan,  Daojin and Gong,  Ming and Guo,  Cheng and Guo,  Shaojun and Han,  Lianchen and Li,  Na and Li,  Shaowei and Li,  Yuan and Liang,  Futian and Lin,  Jin and Qian,  Haoran and Rong,  Hao and Su,  Hong and Sun,  Lihua and Wang,  Shiyu and Wu,  Yulin and Xu,  Yu and Ying,  Chong and Yu,  Jiale and Zha,  Chen and Zhang,  Kaili and Huo,  Yong-Heng and Lu,  Chao-Yang and Peng,  Cheng-Zhi and Zhu,  Xiaobo and Pan,  Jian-Wei},
  year    = {2022},
  month   = jul,
  journal = {Physical Review Letters},
  volume  = {129},
  number  = {3},
  pages   = {030501},
  url     = {http://dx.doi.org/10.1103/PhysRevLett.129.030501},
  doi     = {10.1103/physrevlett.129.030501},
  note    = {clean}
}

@article{acharya2024quantum,
  title   = {Quantum error correction below the surface code threshold},
  author  = {Acharya,  Rajeev and Aghababaie-Beni,  Laleh and Aleiner,  Igor and Andersen,  Trond I. and Ansmann,  Markus and Arute,  Frank and Arya,  Kunal and Asfaw,  Abraham and Astrakhantsev,  Nikita and Atalaya,  Juan and Babbush,  Ryan and Bacon,  Dave and Ballard,  Brian and Bardin,  Joseph C. and Bausch,  Johannes and Bengtsson,  Andreas and Bilmes,  Alexander and Blackwell,  Sam and Boixo,  Sergio and Bortoli,  Gina and Bourassa,  Alexandre and Bovaird,  Jenna and Brill,  Leon and Broughton,  Michael and Browne,  David A. and Buchea,  Brett and Buckley,  Bob B. and Buell,  David A. and Burger,  Tim and Burkett,  Brian and Bushnell,  Nicholas and Cabrera,  Anthony and Campero,  Juan and Chang,  Hung-Shen and Chen,  Yu and Chen,  Zijun and Chiaro,  Ben and Chik,  Desmond and Chou,  Charina and Claes,  Jahan and Cleland,  Agnetta Y. and Cogan,  Josh and Collins,  Roberto and Conner,  Paul and Courtney,  William and Crook,  Alexander L. and Curtin,  Ben and Das,  Sayan and Davies,  Alex and De Lorenzo,  Laura and Debroy,  Dripto M. and Demura,  Sean and Devoret,  Michel and Di Paolo,  Agustin and Donohoe,  Paul and Drozdov,  Ilya and Dunsworth,  Andrew and Earle,  Clint and Edlich,  Thomas and Eickbusch,  Alec and Elbag,  Aviv Moshe and Elzouka,  Mahmoud and Erickson,  Catherine and Faoro,  Lara and Farhi,  Edward and Ferreira,  Vinicius S. and Burgos,  Leslie Flores and Forati,  Ebrahim and Fowler,  Austin G. and Foxen,  Brooks and Ganjam,  Suhas and Garcia,  Gonzalo and Gasca,  Robert and Genois,  Élie and Giang,  William and Gidney,  Craig and Gilboa,  Dar and Gosula,  Raja and Dau,  Alejandro Grajales and Graumann,  Dietrich and Greene,  Alex and Gross,  Jonathan A. and Habegger,  Steve and Hall,  John and Hamilton,  Michael C. and Hansen,  Monica and Harrigan,  Matthew P. and Harrington,  Sean D. and Heras,  Francisco J. H. and Heslin,  Stephen and Heu,  Paula and Higgott,  Oscar and Hill,  Gordon and Hilton,  Jeremy and Holland,  George and Hong,  Sabrina and Huang,  Hsin-Yuan and Huff,  Ashley and Huggins,  William J. and Ioffe,  Lev B. and Isakov,  Sergei V. and Iveland,  Justin and Jeffrey,  Evan and Jiang,  Zhang and Jones,  Cody and Jordan,  Stephen and Joshi,  Chaitali and Juhas,  Pavol and Kafri,  Dvir and Kang,  Hui and Karamlou,  Amir H. and Kechedzhi,  Kostyantyn and Kelly,  Julian and Khaire,  Trupti and Khattar,  Tanuj and Khezri,  Mostafa and Kim,  Seon and Klimov,  Paul V. and Klots,  Andrey R. and Kobrin,  Bryce and Kohli,  Pushmeet and Korotkov,  Alexander N. and Kostritsa,  Fedor and Kothari,  Robin and Kozlovskii,  Borislav and Kreikebaum,  John Mark and Kurilovich,  Vladislav D. and Lacroix,  Nathan and Landhuis,  David and Lange-Dei,  Tiano and Langley,  Brandon W. and Laptev,  Pavel and Lau,  Kim-Ming and Guevel,  Loïck Le and Ledford,  Justin and Lee,  Kenny and Lensky,  Yuri D. and Leon,  Shannon and Lester,  Brian J. and Li,  Wing Yan and Li,  Yin and Lill,  Alexander T. and Liu,  Wayne and Livingston,  William P. and Locharla,  Aditya and Lucero,  Erik and Lundahl,  Daniel and Lunt,  Aaron and Madhuk,  Sid and Malone,  Fionn D. and Maloney,  Ashley and Mandrá,  Salvatore and Martin,  Leigh S. and Martin,  Steven and Martin,  Orion and Maxfield,  Cameron and McClean,  Jarrod R. and McEwen,  Matt and Meeks,  Seneca and Megrant,  Anthony and Mi,  Xiao and Miao,  Kevin C. and Mieszala,  Amanda and Molavi,  Reza and Molina,  Sebastian and Montazeri,  Shirin and Morvan,  Alexis and Movassagh,  Ramis and Mruczkiewicz,  Wojciech and Naaman,  Ofer and Neeley,  Matthew and Neill,  Charles and Nersisyan,  Ani and Neven,  Hartmut and Newman,  Michael and Ng,  Jiun How and Nguyen,  Anthony and Nguyen,  Murray and Ni,  Chia-Hung and O'Brien,  Thomas E. and Oliver,  William D. and Opremcak,  Alex and Ottosson,  Kristoffer and Petukhov,  Andre and Pizzuto,  Alex and Platt,  John and Potter,  Rebecca and Pritchard,  Orion and Pryadko,  Leonid P. and Quintana,  Chris and Ramachandran,  Ganesh and Reagor,  Matthew J. and Rhodes,  David M. and Roberts,  Gabrielle and Rosenberg,  Eliott and Rosenfeld,  Emma and Roushan,  Pedram and Rubin,  Nicholas C. and Saei,  Negar and Sank,  Daniel and Sankaragomathi,  Kannan and Satzinger,  Kevin J. and Schurkus,  Henry F. and Schuster,  Christopher and Senior,  Andrew W. and Shearn,  Michael J. and Shorter,  Aaron and Shutty,  Noah and Shvarts,  Vladimir and Singh,  Shraddha and Sivak,  Volodymyr and Skruzny,  Jindra and Small,  Spencer and Smelyanskiy,  Vadim and Smith,  W. Clarke and Somma,  Rolando D. and Springer,  Sofia and Sterling,  George and Strain,  Doug and Suchard,  Jordan and Szasz,  Aaron and Sztein,  Alex and Thor,  Douglas and Torres,  Alfredo and Torunbalci,  M. Mert and Vaishnav,  Abeer and Vargas,  Justin and Vdovichev,  Sergey and Vidal,  Guifre and Villalonga,  Benjamin and Heidweiller,  Catherine Vollgraff and Waltman,  Steven and Wang,  Shannon X. and Ware,  Brayden and Weber,  Kate and White,  Theodore and Wong,  Kristi and Woo,  Bryan W. K. and Xing,  Cheng and Yao,  Z. Jamie and Yeh,  Ping and Ying,  Bicheng and Yoo,  Juhwan and Yosri,  Noureldin and Young,  Grayson and Zalcman,  Adam and Zhang,  Yaxing and Zhu,  Ningfeng and Zobrist,  Nicholas},
  year    = {2024},
  journal = {arXiv preprint},
  number  = {2408.13687},
  url     = {https://arxiv.org/abs/2408.13687},
  doi     = {10.48550/ARXIV.2408.13687},
  note    = {clean}
}

\clearpage
\appendix

\setcounter{equation}{0}
\renewcommand{\theequation}{S\arabic{equation}}

\setcounter{figure}{0}
\renewcommand{\thefigure}{S\arabic{figure}}

\setcounter{table}{0}
\renewcommand{\thetable}{S\arabic{table}}

\renewcommand{\thesection}{S\arabic{section}}
\renewcommand{\thesubsection}{S\arabic{section}.\arabic{subsection}}
\setcounter{section}{0}

\section{Short derivation of the correlation functions}\label{app:new-proof}

We propose a new derivation of the exact formula for the correlation functions. This derivation has the advantage of being particularly direct from the definition of the SME: by directly applying Itô's lemma we find the exact expression for the generating functional \cref{eq:Zcal}. As explained in \cref{sec:theory}, the correlation functions follow directly from this result.

\textbf{The generating density matrix} -- The main idea is to introduce the \emph{generating density matrix} for a function $j$:
\begin{equation}
    \rho_t^j = \E{\exp\left(\int_{0}^t j_s\,\dd Y_s\right) \rho_t}.
\end{equation}
The generating functional is the trace of the generating density matrix at long time: $\Zcal(j)=\Tr{\rho_\infty^j}$. We will show below that $\rho_t^j$ obeys the simple linear differential equation:
\begin{equation}
    \frac{\dd \rho_t^j}{\dt} = \Lcal^j_t(\rho_t^j),
\end{equation}
where $\Lcal^j_t$ is the superoperator defined by \cref{eq:Lcalj-jump,eq:Lcalj-diff}. In the condensed matter community, this equation is called the \emph{generalised quantum master equation} and $\Lcal^j_t$ is called the \emph{tilted Liouvillian}.

\textbf{Derivation} -- For simplicity, we derive the result for a single detector with jump operator $L$. Let us rewrite:
\begin{equation}
    \rho_t^j = \E{X_t \rho_t}\ \ \text{with}\ \ X_t = \exp\left(\int_{0}^t j_s\,\dd Y_s\right).
\end{equation}
A first application of Itô's lemma for the product rule gives:
\begin{equation}
    \dd \rho_t^j = \E{\dd X_t\,\rho_t + X_t\,\dd\rho_t + \dd X_t\,\dd\rho_t}.\label{eq:ito_rhoj}
\end{equation}
To calculate $\dd X_t$ we apply Itô’s lemma to the exponential. The exact formula depends on the nature of the stochastic process, we consider first the case of the jump SME and then the diffusive SME.

\textit{For the jump SME} -- The signal is $\dd Y_t=\dNt$, and we have:
\begin{equation}
    \dd X_t = (e^{j_t}-1)\dNt\,X_t.
\end{equation}
We then insert this expression into \cref{eq:ito_rhoj} and use the law of $\dNt$ \cref{eq:dNt} and the jump SME \cref{eq:sme,eq:Mjump} for $\dd\rho_t$ to get the result:
\begin{align}
    \frac{\dd \rho_t^j}{\dt} & = \E{X_t
        \Big(
        \Lcal_t(\rho_t) + (e^{j_t}-1) (\theta\rho_t+\eta L\rho_t L^\dag)
    \Big)} \notag                                     \\
                                & = \Lcal^j_t(\rho_t^j),
\end{align}
with $\Lcal^j_t = \Lcal_t + (e^{j_t}-1)\,\Ccal_L$ as in \cref{eq:Lcalj-jump}.

\textit{For the diffusive SME} -- Using \cref{eq:dYt} for the signal $\dd Y_t$ we have:
\begin{equation}
    \dd X_t = \left(j_t \dd Y_t + \frac{j_t^2}{2} \dt\right)X_t.
\end{equation}
We then insert this expression into \cref{eq:ito_rhoj} and use the diffusive SME \cref{eq:sme,eq:Mdiff} for $\dd\rho_t$ to get the result:
\begin{align}
    \frac{\dd \rho_t^j}{\dt} & = \E{X_t
        \Big(
        \Lcal_t(\rho_t) + j_t\sqrt{\eta} (L\rho_t+\rho_t L^\dag) + \frac{j_t^2}{2}\rho_t
    \Big)} \notag                                     \\
                                & = \Lcal^j_t(\rho_t^j),
\end{align}
with $\Lcal^j_t = \Lcal_t + j_t\,\Ccal_L+j_t^2/2$ as in \cref{eq:Lcalj-diff}.

We recommend reading our earlier work \cite{guilmin2023correlation} for a more intuitive, albeit longer, derivation of the result.

\section{System of ODEs for filtered signal correlation functions}\label{app:filtered}

For convenience, we give the system of ODEs to solve to compute correlation functions up to order three for a single signal. Note that these formulae are straightforward to find back by manually forward differentiating the ODE \cref{eq:ODE} as explained in the main text. We drop the time index for compactness.

\textbf{One-point correlation function} -- For the jump or diffusive SME:
\begin{equation}
    \begin{bmatrix} \dot\rho\\ \dot\rho^1\end{bmatrix}=
    \begin{bmatrix}
        \Lcal      & 0     \\
        f_1\Ccal_L & \Lcal
    \end{bmatrix}
    \begin{bmatrix} \rho\\ \rho^1\end{bmatrix}
\end{equation}

\textbf{Two-point correlation function} -- For the jump SME:
\begin{equation}
    \begin{bmatrix} \dot\rho\\ \dot\rho^1\\ \dot\rho^2\\ \dot\rho^{12}\end{bmatrix}=
    \begin{bmatrix}
        \Lcal           & 0            & 0            & 0     \\
        f_1\Ccal_L      & \Lcal        & 0            & 0     \\
        f_2\,\Ccal_L    & 0            & \Lcal        & 0     \\
        f_1f_2\,\Ccal_L & f_2\,\Ccal_L & f_1\,\Ccal_L & \Lcal
    \end{bmatrix}
    \begin{bmatrix} \rho\\ \rho^1\\ \rho^2\\ \rho^{12}\end{bmatrix}
\end{equation}
For the diffusive SME (\cref{eq:f2-correlation} in the main text):
\begin{equation}
    \begin{bmatrix} \dot\rho\\ \dot\rho^1\\ \dot\rho^2\\ \dot\rho^{12}\end{bmatrix}=
    \begin{bmatrix}
        \Lcal        & 0            & 0            & 0     \\
        f_1\Ccal_L   & \Lcal        & 0            & 0     \\
        f_2\,\Ccal_L & 0            & \Lcal        & 0     \\
        f_1f_2       & f_2\,\Ccal_L & f_1\,\Ccal_L & \Lcal
    \end{bmatrix}
    \begin{bmatrix} \rho\\ \rho^1\\ \rho^2\\ \rho^{12}\end{bmatrix}
\end{equation}

\textbf{Three-point correlation function}
See \cref{eq:jump-3pt} for the jump SME and \cref{eq:diff-3pt} for the diffusive SME.

\begin{figure*}[ht]
For the jump SME:
\begin{equation}\label{eq:jump-3pt}
    \begin{bmatrix} \dot\rho\\ \dot\rho^1\\ \dot\rho^2\\ \dot\rho^3\\ \dot\rho^{12} \\ \dot\rho^{13} \\ \dot\rho^{23} \\ \dot\rho^{123}\end{bmatrix}=
    \begin{bmatrix}
        \Lcal & 0     & 0 & 0 & 0 & 0 & 0 & 0 \\
        f_1\,\Ccal_L & \Lcal & 0 & 0 & 0 & 0 & 0 & 0 \\
        f_2\,\Ccal_L     & 0     & \Lcal & 0 & 0 & 0 & 0 & 0 \\
        f_3\,\Ccal_L     & 0     & 0 & \Lcal & 0 & 0 & 0 & 0 \\
        f_1f_2\,\Ccal_L     & f_2\,\Ccal_L     & f_1\,\Ccal_L & 0 & \Lcal & 0 & 0 & 0 \\
        f_1f_3\,\Ccal_L     & f_3\,\Ccal_L     & 0 & f_1\,\Ccal_L & 0 & \Lcal & 0 & 0 \\
        f_2f_3\,\Ccal_L     & 0     & f_3\,\Ccal_L & f_2\,\Ccal_L & 0 & 0 & \Lcal & 0 \\
        f_1f_2f_3\,\Ccal_L     & f_2f_3\,\Ccal_L     & f_1f_3\,\Ccal_L & f_1f_2\,\Ccal_L & f_3\,\Ccal_L & f_2\,\Ccal_L & f_1\,\Ccal_L & \Lcal
    \end{bmatrix}
    \begin{bmatrix} \rho\\ \rho^1\\ \rho^2\\ \rho^3\\ \rho^{12} \\ \rho^{13} \\ \rho^{23} \\ \rho^{123}\end{bmatrix}
\end{equation}

For the diffusive SME:
\begin{equation}\label{eq:diff-3pt}
    \begin{bmatrix} \dot\rho\\ \dot\rho^1\\ \dot\rho^2\\ \dot\rho^3\\ \dot\rho^{12} \\ \dot\rho^{13} \\ \dot\rho^{23} \\ \dot\rho^{123}\end{bmatrix}=
    \begin{bmatrix}
        \Lcal        & 0            & 0            & 0            & 0            & 0            & 0            & 0     \\
        f_1\,\Ccal_L & \Lcal        & 0            & 0            & 0            & 0            & 0            & 0     \\
        f_2\,\Ccal_L & 0            & \Lcal        & 0            & 0            & 0            & 0            & 0     \\
        f_3\,\Ccal_L & 0            & 0            & \Lcal        & 0            & 0            & 0            & 0     \\
        f_1f_2       & f_2\,\Ccal_L & f_1\,\Ccal_L & 0            & \Lcal        & 0            & 0            & 0     \\
        f_1f_3       & f_3\,\Ccal_L & 0            & f_1\,\Ccal_L & 0            & \Lcal        & 0            & 0     \\
        f_2f_3       & 0            & f_3\,\Ccal_L & f_2\,\Ccal_L & 0            & 0            & \Lcal        & 0     \\
        0            & f_2f_3       & f_1f_3       & f_1f_2       & f_3\,\Ccal_L & f_2\,\Ccal_L & f_1\,\Ccal_L & \Lcal
    \end{bmatrix}
    \begin{bmatrix} \rho\\ \rho^1\\ \rho^2\\ \rho^3\\ \rho^{12} \\ \rho^{13} \\ \rho^{23} \\ \rho^{123}\end{bmatrix}
\end{equation}
\end{figure*}

\section{Numerical considerations}\label{app:numerical}

\subsection{Exploiting the generator block structure}

The generator of the ODE system to solve for the filtered and binned signal, e.g. \cref{eq:f2-correlation} for the two-point function, is naturally in block form. This can be exploited to reduce the numerical cost of computing filtered and binned correlation functions to that of solving a single ME, by vectorising the application of the generator to the states. Note that this is only valid for low-order correlation functions, as the number of states scales as $2^n$ with $n$ the number of points.

\subsection{Vectorising the computation of multiple correlation functions}

The computation can often be efficiently vectorised to compute multiple correlation functions at different set of times. For example using \cref{eq:f2-correlation} to compute the binned signal two-point correlation functions $\E{I_0I_k}$ for different time bins $1\leq k\leq N$, the states $\rho^2$ and $\rho^{12}$ remain null before time $t_k=k\Delta t$. The quantities $\E{I_0I_k}$ can then be computed in a vectorised way by (i) evolving the states $(\rho_t,\rho_t^1)$ on $[0,t_1)$ with the one-point system, then (ii) evolving the resulting states $(\rho_t,\rho_t^1)$ with the Liouvillian on $[t_1, t_N)$, and saving the result $(\rho_{t_k},\rho^1_{t_k})$ at intermediate times $t_k$, and finally (iii) computing the evolution of the $N$ states $(\rho_t,\rho_t^1,\rho_t^2,\rho_t^{12})_k$ on $[t_k,t_{k+1})$ in a vectorised fashion using the two-point system, initialised with the stored results $(\rho_{t_k},\rho^1_{t_k}, 0, 0)$ from the previous step.

\subsection{Normalising the generator for adaptive step-size ODE solvers}

When solving the ODE system with an adaptive step-size solver, special care must be taken to ensure that the non-zero generator elements are of the same order of magnitude. Taking for example \cref{eq:f2-correlation}, if the terms $f_1\,\Ccal_L$, $f_2\,\Ccal_L$ or $f_1f_2$ differ significantly in magnitude compared to the term $\Lcal$, the error estimate of the ODE solver used to adaptively adjust the step size might be significantly under- or over-estimated. In practice, underestimation leads to excessively large time steps and thus erroneous results, while overestimation leads to excessively small time steps and thus prolonged solving time. The latter situation typically occurs for small time bins $\Delta t\ll 1$, for example for binned signals with ${f_k = 1/\Delta t\ \1_{[k\Delta t, (k+1)\Delta t)}}$, the resulting function is very large, $f_k(t)\gg 1$. A simple numerical trick is to normalise the generator by scaling the filter functions $f_k$ by a constant factor $g$: $\tilde{f_k}=f_k/g$. Assuming that the Lindbladian timescales have already been normalised, one can choose $g$ such that $\tilde f_k$ is of the order of unity. The correlation functions $\mathbb{E}[I_{\tilde{f}_1}\dots I_{\tilde{f}_n}]$ can then be computed with a regular adaptive step-size ODE solver, and the correct correlation functions are found by simply scaling back the result: ${\mathbb{E}[I_{f_1}\dots I_{f_n}] = g^n\, \mathbb{E}[I_{\tilde{f}_1}\dots I_{\tilde{f}_n}]}$.

\section{Experimental considerations}\label{app:experimental}

\subsection{Estimating uncertainties of experimental correlation functions}\label{app:uncertainties}

We recall that the experimental estimate of the correlation function for the $\nexp$ realisation of an experiment is (\cref{eq:Ehat} in the main text):
\begin{equation}
    \Ehat{I_{k_1}\dots I_{k_n}} = \frac{1}{\nexp}\sum_{j=1}^\nexp I_{k_1}^{(j)}  \dots I_{k_n}^{(j)},
\end{equation}
where $\{I_0^{(j)},\dots,I_N^{(j)}\}$ is the digitised signal recorded during the $j$-th experiment.

The uncertainty of this experimental estimate is given by the standard error of the mean (SEM) $\sigma_{\hat{\mathbb{E}}[I_{k_1}\dots I_{k_n}]}$, which quantifies how much the sample mean $\Ehat{\dots}$ is expected to deviate from the true mean $\E{\dots}$. The SEM itself can be estimated from the data, by computing ${\hat{\sigma}_{\hat{\mathbb{E}}[I_{k_1}\dots I_{k_n}]} = \hat{\sigma} / \sqrt\nexp}$ where $\hat{\sigma}^2$ is the unbiased sample variance:
\begin{equation}\notag
    \hat{\sigma}^2 = \frac{1}{\nexp-1}\sum_{j=1}^\nexp\left(I_{k_1}^{(j)} \dots I_{k_n}^{(j)}-\Ehat{I_{k_1}\dots I_{k_n}}\right)^2.
\end{equation}
The number of realisations needed to improve the precision of the estimate $\Ehat{I_{k_1}\dots I_{k_n}}$ scales as $\sqrt\nexp$, and the prefactor depends on the measurement efficiency.

Note that for a given number of experiments $\nexp$, estimates of higher-order correlation functions typically have a much larger standard error of the mean (SEM), and thus accurate estimation of higher-order correlation functions quickly becomes prohibitively expensive in measurement time.

\subsection{Correlation functions in the steady state}\label{app:steady-state}

When starting from the system steady state, the measured signal is a stationary stochastic process: shifting the entire process by some time $\delta$ does not change its statistical properties. Thus, a single long trajectory can be measured, instead of repeating the experiment multiple times. The measured signal can then be divided into $\nexp$ chunks, and the average is computed by considering each chunk as a different realisation of the experiment. Certain technical considerations must be taken into account for this approach to be rigorous: (i) the steady state must be ergodic under continuous measurement, and (ii) there must be sufficient statistical independence between the different chunks (which is generally ensured by the fact that the correlation functions decay to zero at long times).

In this situation, the one-point correlation function reduces to a single value: $\E{I_t}=\E{I_0}$, and the two-point correlation functions $\E{I_{t_1}I_{t_2}}$ depend on a single parameter $\tau=t_2-t_1$: ${\E{I_{t_1}I_{t_2}}=\E{I_{0}I_{t_2-t_1}}=\E{I_{0}I_\tau}}$. More generally, for any times $t_1, \dots, t_n$ and any time shift $\delta$ we have ${\E{I_{t_1} I_{t_2} \dots I_{t_n}} = \E{I_{t_1+\delta} I_{t_2+\delta} \dots I_{t_n+\delta}}}$.

\subsection{Choosing the digitisation time}\label{app:digitisation}

The choice of digitisation time $\Delta t$ is a compromise between choosing a small time to capture the dynamics of the system without averaging important timescales, and the technical constraints that force the experimentalist to choose a large time. These constraints include the maximum sampling rate of the readout electronics, the need to reduce high frequency noise in certain situations, and the memory and computational cost associated with processing a very finely sampled signal. To give an order of magnitude, for heterodyne detection using a modern analogue-to-digital converter (ADC) with a sampling rate of $1\ \text{GHz}$ (${\Delta t=1\ \text{ns}}$), one second of signal stored in simple precision (\texttt{float32}) represents about $7.5\ \text{GB}$ of data. This is particularly constraining for long measurements, multiple signals or real time applications. The digitisation time must thus be chosen small enough to capture the essential dynamics of the system while respecting the experimental constraints.

As explained in the main text, if the digitisation time $\Delta t$ is very small compared to the fastest timescale of the system, it is tempting to approximate the binned signal correlation functions using the sharp signal formula \cref{eq:sharp-correlation}. This approximation is not always valid, and the resulting error should be estimated beforehand in simulation by comparing the result with the exact binned signal formula. As an illustration, for the first example (\cref{sec:ex1}), even though the time bin $\Delta t$ is relatively small compared to the system timescale, using this approximation results in relative errors on the estimated parameters that are more than an order of magnitude larger than those reported in the main text. Importantly, this error is not a statistical error that can be reduced by averaging over more trajectories, but a constant bias due to the incorrect approximation.

\section{Parameters degeneracy}\label{app:degeneracy}

In this section, we discuss a number of potential degeneracies that we have observed in practice. We also explain how to identify them, and possible solutions to lift these degeneracies.

\textbf{Fundamental non-identifiability} -- In some situations, the parameters are not identifiable in the measured signal. As a simple example, consider a lossy harmonic oscillator with Hamiltonian and jump operator:
\begin{align}\label{eq:oscillator}
    \begin{split}
        H & = \omega a^\dag a, \\
        L & = \sqrt{\kappa} a,
    \end{split}
\end{align}
where $a$ is the oscillator annihilation operator, $\omega$ is the oscillator frequency and $\kappa$ is the single-photon loss rate. The loss channel is monitored with efficiency $\eta$ by homodyne detection along the $X$ quadrature. Suppose the system starts from a coherent state $\rho_0=\ket{\alpha_0}\bra{\alpha_0}$ and our goal is to find the values of the four parameters $\btheta=(\omega, \kappa, \eta, \alpha_0)$. Then it is straightforward to see that the parameters $\eta$ and $\alpha_0$ cannot be identified from the measured signal. Indeed, the state remains a coherent state during the evolution: $\rho_t=\ket{\alpha_t}\bra{\alpha_t}$ with $\alpha_t=\alpha_0 e^{-\kappa/2 t}e^{-i\omega t}$, because the measurement backaction is null at all time: $\Mcal(\rho_t,\dYt)=0$. The measured signal is then:
\begin{align}
    \dYt &= \sqrt{\eta\kappa}\mathrm{Tr}\left[(a+a^\dag)\rho_t\right]\dt + \dWt\\
            &= 2\sqrt{\eta\kappa}\alpha_0 e^{-\kappa/2 t}\cos(\omega t)\,\dt + \dWt,
\end{align}
which allows $\kappa$ and $\omega$ to be uniquely identified, but not $\eta$ and $\alpha_0$, as they only appear as the product $\sqrt\eta \alpha_0$.

\textbf{Low-order degeneracy} -- Sometimes, the chosen set of correlation functions is not sufficient to estimate all the parameters. This is illustrated in the third example (\cref{sec:ex3}), where the odd-order correlation functions are all null, and the two-point correlation functions are degenerate: different combinations of parameters give the same correlation functions.

\textbf{Statistical degeneracy} -- We could also refer to this situation as ``practical'' degeneracy: while the chosen correlation functions are sufficient to estimate all the parameters, in practice, a very large number of trajectories are needed to achieve low error bars on the estimated parameters.

\textbf{Possible solutions} -- A first general solution to lift these degeneracies is to change the dynamics. In the simple oscillator example \cref{eq:oscillator}, the degeneracy can be lifted by adding an unknown linear drive, or by making the oscillator anharmonic, even if this requires increasing the number of parameters to fit.

Another general solution is to measure the system in multiple configurations and jointly fit the correlation functions, for example starting from different initial states, or by scaling a controllable parameter by a known factor, such as a linear drive. This solution was illustrated in the third example (\cref{sec:ex3}), where measuring the system for different values of $\alpha_2$ lifts the observed degeneracy. In practice, this idea can be taken further by systematically varying every available control knob to measure correlation functions across different regimes of the dynamics. This approach is particularly effective for ensuring that there are no degeneracies, and for fitting the parameters with high precision.

A specific solution to the low-order degeneracy is to resort to higher-order correlation functions, as was also illustrated in the third example (\cref{sec:ex3}), where the four-point correlation functions lift the degeneracy. This can also help in cases of a statistical degeneracy: sometimes, certain parameters are easier to identify with higher-order correlation functions.

\textbf{Identify parameter degeneracy} -- For a given system, there are simple ways to determine whether there is a degeneracy, before doing the experiment and without having to simulate SME trajectories. To verify that there is no fundamental non-identifiability or low-order degeneracy, we check whether we can directly fit the exact correlation functions (computed numerically for the true parameter values, with no statistical error). In all proposed examples, we recover the true parameters without error. This ensures that, for the chosen set of parameters and correlation functions, the fit should converge if we have a sufficient number of trajectories.

A practical way to check the last type of degeneracy, statistical degeneracy, is to artificially add some noise -- for example Gaussian sampled noise --, to the exact correlation functions computed numerically. One can then verify whether the fit to this noisy data converges close to the true parameters.

If numerical differentiation of the least-squares loss function is available, there is a more quantitative alternative. The Hessian of the loss function quantifies how sensitive it is to a small change in each parameter, and can help to identify potential parameter degeneracy. This numerical study addresses whether a set of parameters can, in principle, be identified by a given experiment.

\section{Estimating the gain}\label{app:gain}

For diffusive measurements, the total gain $G$ of the amplification chain is typically unknown. If the system remains in vacuum at all time $\rho_t=\ket0\bra0$, we have $\dd Y_t=\dd W_t$ and the measured signal $I_k$ is just filtered white noise, with autocorrelation:
\begin{align}
    \E{I_k^2} & = \E{\left(\frac{G}{\Delta t}\int_{k\Delta t}^{(k+1)\Delta t}\,\dWt\right)^2}             \\
                & = \E{\frac{G^2}{(\Delta t)^2}\int_{k\Delta t}^{(k+1)\Delta t}\dt} = \frac{G^2}{\Delta t},
\end{align}
where we used Itô isometry for the passage to the second line. Thus, the gain $G$ can be estimated directly from the autocorrelation of the signal when the system remains in vacuum. Note that this characterisation of the gain does not require knowing the efficiency $\eta$.

\section{Details on each example}

For each example, we ran the procedure for various initial guesses, both close to and far from the true parameters. We observed either convergence to the true parameters or to a local minimum with a visually poor goodness of fit, which can be easily disregarded.

\subsection{Heterodyne measurement of a driven anharmonic oscillator}\label{app:ex1}

\begin{figure}[htb]
    \centering
    \includegraphics{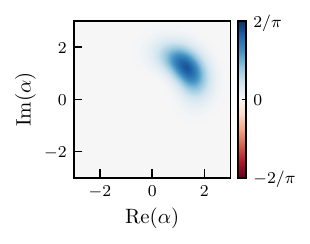}
    \vspace{-0.4cm}
    \caption{\textbf{Wigner function of the driven anharmonic oscillator steady state.}}
    \label{fig:ex1/wigner}
\end{figure}

\textbf{Initial state} -- The system is initialised in its steady state, whose Wigner function is shown in \cref{fig:ex1/wigner}.

\begin{figure*}[htb]
    \includegraphics{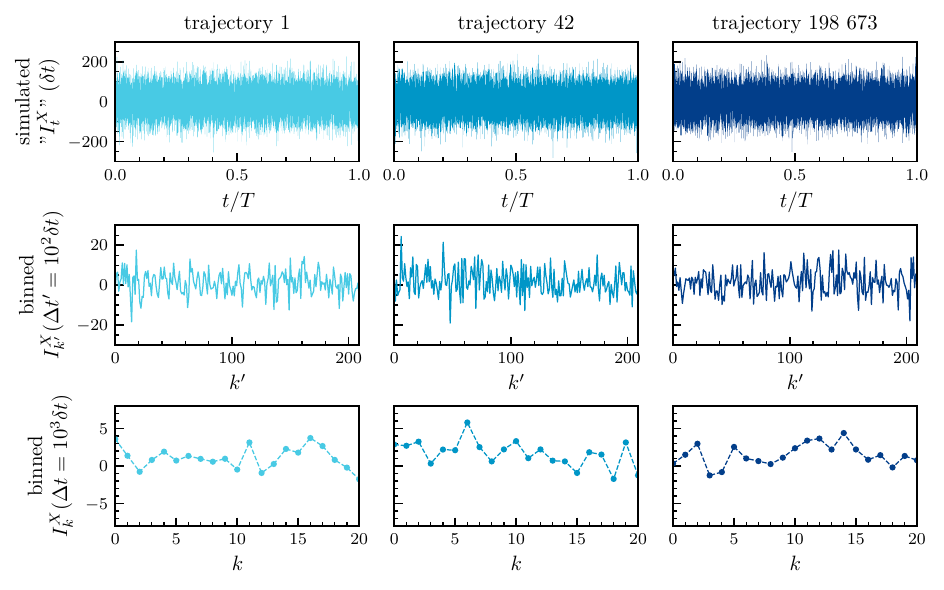}
    \vspace{-0.4cm}
    \caption{\textbf{Simulated diffusive signal.} Three different trajectories (left, middle and right columns) for the heterodyne measurement of the $X$ quadrature of the driven anharmonic oscillator are simulated from time $0$ to time $T=21\Delta t$ by starting from a different random seed. The first row shows the simulated ``sharp'' signal for the fine-grained time step $\delta t$, and the second and third rows show the digitised signal after averaging for durations of $\Delta t'=100\,\delta t$ and $\Delta t=1000\,\delta t$ (main text value).}
    \label{fig:ex1/trajectories}
\end{figure*}

\textbf{Trajectories simulation} -- For the simulation, we chose a Fock space truncation of $16$, and a fixed step size for the numerical scheme $\delta t=\Delta t/1000$, to ensure numerical accuracy (where $\Delta t$ is the duration of a single time bin, introduced in the main text). An example of simulated trajectories before and after binning is shown in \cref{fig:ex1/trajectories}.

\textbf{Initial guess} -- The parameters guess to initialise the fit in the main text is given in \cref{table:ex1-guess}.

\begin{table}[htb]
    \centering
    \begin{tabular}{|c|l|l|l|}
        \hline
                & \textbf{Value}      & \textbf{Guess}\\\hline\hline
        $K / (2\pi)$                   & $100\ \mathrm{kHz}$ & $150\ \mathrm{kHz}$ \\
        $\epsilon_x / (2\pi)$        & $300\ \mathrm{kHz}$ & $200\ \mathrm{kHz}$\\
        $\epsilon_y / (2\pi)$   & $400\ \mathrm{kHz}$ & $200\ \mathrm{kHz}$\\
        $\eta$                & $0.8$               & $0.6$              \\\hline
    \end{tabular}
    \caption{\textbf{Initial parameters guess for the driven anharmonic oscillator.}}
    \label{table:ex1-guess}
\end{table}

\textbf{Other correlation functions} -- Note that we could also have included other correlation functions in the fit: the mean values $\E{I_0^X}$ and $\E{I_0^P}$, the coincident two-point correlation functions for each signal $\E{I_0^XI_0^X}$ and $\E{I_0^PI_0^P}$, the two-point correlation functions between the two signals $\E{I_{k_1}^XI_{k_2}^P}$, and any higher-order correlation functions. However, we didn't deem it necessary for this system, as the fit was satisfactory.

\textbf{Fitting $\kappa$ as well} -- We did not include the decay rate $\kappa$ in the unknown parameters because it is usually estimated with good precision from a basic spectroscopic measurement. However, it is possible to also fit $\kappa$ simultaneously with the other parameters.

\subsection{Jump measurement of a driven two-level system} \label{app:ex2}

\begin{figure}[htb]
    \centering
    \includegraphics{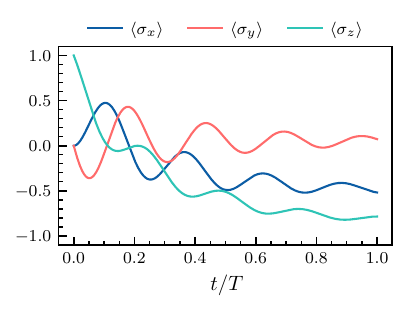}
    \vspace{-0.4cm}
    \caption{\textbf{Average Lindblad evolution of the driven two-level system.} ME simulation from time $0$ to time $T=21\Delta t$ of the three Pauli observables $\sigma_x$, $\sigma_y$ and $\sigma_z$ expectation values.}
    \label{fig:ex2/lindblad}
\end{figure}

\textbf{Average dynamics} -- The system average dynamics when starting from the excited state $\rho_0=\ket1\bra1$ is given by the Lindblad evolution shown in \cref{fig:ex2/lindblad}.

\begin{figure*}[htb]
    \includegraphics{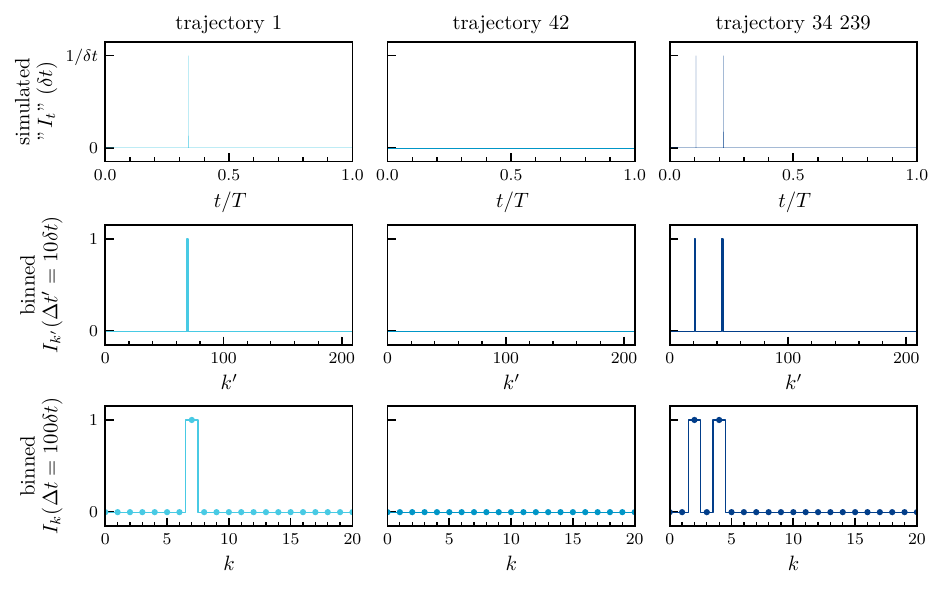}
    \vspace{-0.4cm}
    \caption{\textbf{Simulated jump signal.} Three different trajectories (left, middle and right columns) for the jump measurement of the driven two-level system are simulated from time $0$ to time $T=21\Delta t$ by starting from a different random seed. The first row shows the simulated ``sharp'' signal for the fine-grained time step $\delta t$, and the second and third rows show the digitised signal after averaging for durations of $\Delta t'=10\,\delta t$ and $\Delta t=100\,\delta t$ (main text value).}
    \label{fig:ex2/trajectories}
\end{figure*}

\textbf{Trajectories simulation} -- For the simulation, we chose a fixed step size for the numerical scheme $\delta t=\Delta t/100$, to ensure numerical accuracy (where $\Delta t$ is the duration of a single time bin, introduced in the main text). An example of simulated trajectories, before and after binning is shown in \cref{fig:ex2/trajectories}.

\textbf{Initial guess} -- The parameters guess to initialise the fit in the main text is given in \cref{table:ex2-guess}.

\begin{table}[htb]
    \centering
    \begin{tabular}{|c|l|l|l|}
        \hline
                & \textbf{Value}      & \textbf{Guess}\\\hline\hline
        $\Delta/(2\pi)$                   & $5\ \mathrm{kHz}$ & $4\ \mathrm{kHz}$ \\
        $\Omega/(2\pi)$        & $3\ \mathrm{kHz}$ & $4\ \mathrm{kHz}$\\
        $\gamma/(2\pi)$   & $2\ \mathrm{kHz}$ & $1\ \mathrm{kHz}$\\
        $\theta/(2\pi)$                & $300\ \mathrm{Hz}$               & $500\ \mathrm{Hz}$\\
        $\eta$                & $0.5$               & $0.7$              \\\hline
    \end{tabular}
    \caption{\textbf{Initial parameters guess for the driven two-level system.}}
    \label{table:ex2-guess}
\end{table}

\subsection{Homodyne measurement of a two-photon dissipative oscillator}\label{app:ex3}

\begin{figure}[htb]
    \centering
    \includegraphics{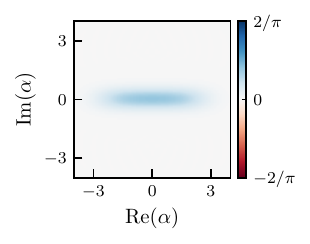}
    \vspace{-0.4cm}
    \caption{\textbf{Wigner function of the two-photon dissipative oscillator steady state.}}
    \label{fig:ex3/wigner}
\end{figure}

\textbf{Initial state} -- The system is initialised in its steady state, whose Wigner function is shown in \cref{fig:ex3/wigner}.

\textbf{Trajectories simulation} -- For the simulation, we chose a Fock space truncation of $32$, and a fixed step size for the numerical scheme $\delta t=\Delta t/500$, to ensure numerical accuracy (where $\Delta t$ is the duration of a single time bin, introduced in the main text).

\textbf{Initial guess} -- The parameters guess to initialise the fit in the main text is given in \cref{table:ex3-guess}.

\begin{table}[htb]
    \centering
    \begin{tabular}{|c|l|l|l|}
        \hline
                & \textbf{Value}      & \textbf{Guess}\\\hline\hline
        $\kappa_2/(2\pi)$        & $1\ \mathrm{kHz}$ & $1.5\ \mathrm{kHz}$\\
        $\alpha_2$   & $7$ & $6$\\
        $\eta$                & $0.1$               & $0.05$              \\\hline
    \end{tabular}
    \caption{\textbf{Initial parameters guess for the two-photon dissipative oscillator.}}
    \label{table:ex3-guess}
\end{table}

\section{Diffusive continuous measurement and system symmetry}\label{app:symmetry}

In this appendix we explain how certain symmetries of a quantum system are reflected in the measured signal. We give a sufficient set of conditions for the diffusive measurement signal $I_t$ to be a symmetric stochastic process. These conditions are valid for a generic quantum system, under any diffusive continuous measurement. We give some intuition for the result, and we apply it to the system studied in \cref{sec:ex3}, to explain why all odd-order correlation functions of the measured signal are null.

We say that a stochastic process is \emph{symmetric} if its distribution is symmetric about the origin, or equivalently if all its odd-order correlation functions are null. For simplicity, we consider the signal $I_t$ from a single diffusive detector monitoring the loss channel $L$, and a constant Liouvillian $\Lcal_t=\Lcal$. The result is easily generalised to multiple detectors and for a time-dependent Liouvillian.

We consider symmetries with respect to a trace-preserving superoperator $\Scal$, which we call a \textit{symmetry superoperator}. An operator $O$ is \textit{symmetric} with respect to $\Scal$ if $O=\Scal(O)$, it is \textit{antisymmetric} w.r.t. $\Scal$ if $O=-\Scal(O)$.

\textbf{Theorem} -- If the following conditions are met:
\begin{enumerate}[(i)]
    \item the Liouvillian commutes with a symmetry superoperator: $[\Lcal,\Scal]=0$,
    \item the initial quantum state $\rho_0$ is symmetric with respect to the symmetry superoperator $\Scal$: $\rho_0 = \Scal(\rho_0)$,
    \item the diffusive correlation superoperator $\Ccal_L$ given by \cref{eq:Lcalj-diff} anticommutes with the symmetry superoperator: $\{\Ccal_L,\Scal\}=0$ (i.e. for any operator $O$, $\Ccal_L\Scal(O)=-\Scal \Ccal_L(O)$),
\end{enumerate}
then the measured signal $I_t$ is a \emph{symmetric} stochastic process.

\textbf{Proof} -- Let us recall the formula for the correlation functions of the sharp signal \cref{eq:sharp-correlation} for $t_1<\dots<t_n$ and a constant Liouvillian:
\begin{equation}
    \E{I_{t_1}\dots I_{t_n}} = \Tr{\Ccal_L e^{(t_n-t_{n-1})\Lcal} \dots \Ccal_L e^{t_1\Lcal} (\rho_0)}.\notag
\end{equation}
The core of the proof lies in the following observations:
\begin{enumerate}
    \item The ensemble-averaged evolution $e^{(t_k-t_{k-1})\Lcal}$ preserves an operator symmetry or anti-symmetry. Indeed, if the superoperators $\Lcal$ and $\Scal$ commute, so does the evolution superoperator $e^{t\Lcal}$, and as a consequence evolutions generated by the Liouvillian preserve the original symmetry of an operator $O$: ${\Scal e^{t\Lcal}(O) = e^{t\Lcal}\Scal(O) = e^{t\Lcal}(O)}$, so $e^{t\Lcal}(O)$ is symmetric.
    \item The inserted correlation superoperator $\Ccal_L$ at each time $t_k$ switches an operator symmetry.
    \item The trace of an antisymmetric operator is null. If ${O=-\Scal(O)}$ then $\Tr{O}=-\Tr{\Scal(O)}=-\Tr{O}$ (because $\Scal$ is trace-preserving), so $\Tr{O}=0$.
\end{enumerate}

Thus, starting from a symmetric operator $\rho_0$, an odd-order correlation function switches the symmetry an odd number of times. The resulting operator is antisymmetric so its trace is null, resulting in a null correlation function, which completes the proof.

Note that for binned signals, the odd-order correlation functions are null if the bins do not overlap, and more generally for filtered signals, the odd-order correlation functions are null for any filter function with empty intersecting support.

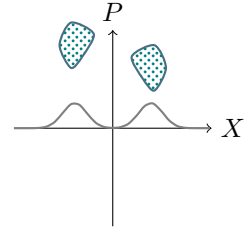
\begin{figure}[htb]
    \centering
    \begin{tikzpicture}
        \draw[->] (-1.3,0) -- (1.3,0) node[right] {$X$};
        \draw[->] (0,-1.3) -- (0,1.3) node[above] {$P$};
        \potato{(-0.7, 1.1)}{0}{(-0.9, 1.1)}{}{1};
        \potato{(0.7, 0.8)}{0}{(1.2, 0.8)}{}{-1};
        \draw[smooth, thick, domain=-1.3:1.28, gray] plot (\x, {exp(-(\x+0.5)*(\x+0.5)/0.05)/3 + exp(-(\x-0.5)*(\x-0.5)/0.05)/3});
    \end{tikzpicture}
    \caption{\textbf{An intuitive illustration of the connection between symmetric dynamics and symmetric measurement.} Sketch of a symmetric state (blue potatoes) for the $X$ quadrature continuous measurement (grey curve).}
    \label{fig:symmetry}
\end{figure}

\textbf{Intuition for the result} -- Consider a bosonic mode whose $X$ quadrature is measured continuously. Intuitively, if the system has some symmetry w.r.t. this quadrature, then the measured signal will be symmetric. This is illustrated in \cref{fig:symmetry}.

Note that under continuous monitoring, the quantum state \emph{does not remain symmetric}, due to the measurement backaction: along a single quantum trajectory, the measurement of the $X$ quadrature breaks the symmetry. Statistically, however, the average over all possible trajectories will inherit from the symmetry of the dynamics.

\textbf{Application to the two-photon dissipative oscillator} -- As we show below, the two-photon dissipation and the single-photon loss \cref{eq:HL1L2} preserve a state symmetry under a rotation of angle $\pi$. Consequently, monitoring any quadrature leads to a symmetric stochastic signal.

\begin{figure}[htb]
    \centering
    \begin{tikzpicture}
        \draw[->] (-1.3,0) -- (1.3,0) node[right] {$X$};
        \draw[->] (0,-1.3) -- (0,1.3) node[above] {$P$};
        \potato{(-0.7, 0.8)}{0}{(-0.9, 0.8)}{$\rho$}{1};
        \potato{(0.7, -0.8)}{180}{(1.2, -0.9)}{$\Pcal(\rho)$}{1};
        \draw[->,>=stealth] (-0.55,0.35) arc (0:75:-1.0) node[midway,xshift=-4,yshift=-4] {$\pi$};
    \end{tikzpicture}
    \caption{\textbf{Sketch of the action of the parity superoperator.} The parity superoperator $\Pcal$ acts on a state $\rho$ in phase space by rotating it by an angle $\pi$.}
    \label{fig:pi-rotation}
\end{figure}
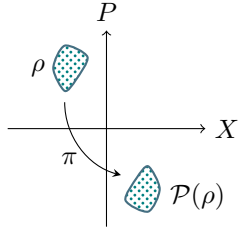

First, we define the \emph{parity superoperator} $\Pcal$ that rotates a state by an angle $\pi$ in phase space: ${\Pcal(\bullet)=e^{i\pi a^\dag a}\bullet e^{-i\pi a^\dag a}}$, as illustrated in \cref{fig:pi-rotation}. A state is symmetric w.r.t. $\Pcal$ if it is unchanged by a rotation of the angle $\pi$ in phase space.

Then we can check the three conditions for the theorem to hold.
\begin{enumerate}[(i)]
    \item The Liouvillian commutes with the symmetry superoperator $[\Lcal,\Pcal]=0$. This has already been observed in \cite{lieu2020symmetry} where it is called a \textit{weak symmetry} of the system.
    \item  We consider the correlation functions of the signal once the steady state $\rho_\infty$ has been reached: $\Lcal(\rho_\infty)$. This system has a unique steady state \cite{minganti2016exact}. The state $\rho_\infty$ is symmetric w.r.t. $\Pcal$, because $\Pcal(\rho_\infty)$ is also a steady state ($\Lcal\Pcal(\rho_\infty) = \Pcal\Lcal(\rho_\infty) = 0$), so by uniqueness of the steady state $\rho_\infty=\Pcal(\rho_\infty)$.
    \item The correlation superoperator $\Ccal_L$ with $L=\sqrt{\kappa}e^{i\varphi} a$ for the quadrature of phase $\varphi\in[0, 2\pi)$ anticommutes with the parity superoperator.
\end{enumerate}
The measured signal is therefore a symmetric stochastic process.

Note that the result also holds if we add a self-Kerr term in the Hamiltonian $H=-Ka^{\dag2}a^2/2$ and/or a dephasing term $\Dcal[\sqrt{2\kappa_\varphi}a^\dag a]$, since the resulting Liouvillian still commutes with the parity superoperator. More generally, any additional term that preserves this commutation relation preserves the symmetry. For example, the result also holds if such a stabilisation is engineered using a so-called \emph{buffer} mode with annihiliation operator $b$, where the Hamiltonian is $H=g_2^*a^2 b^\dag+g_2a^{\dag 2} b+\epsilon^* b+\epsilon b^\dag$ and with loss operators $L_a=\sqrt{\kappa_a}a$ and $L_b=\sqrt{\kappa_b}b$, when any quadratures of mode $a$ and/or mode $b$ are monitored.

\end{document}